\documentclass[prd,onecolumn,superscriptaddress,amsfonts,amssymb,amsmath,showpacs]{revtex4-2}
\usepackage{bm}
\usepackage{latexsym}
\usepackage[latin1]{inputenc}
\usepackage{graphicx}
\usepackage{amsmath}
\usepackage{amssymb}
\usepackage{amsthm}
\usepackage{relsize}
\usepackage{palatino}
\usepackage{mathpazo}
\usepackage{textcomp}
\usepackage{float}
\usepackage{booktabs}
\usepackage{dcolumn}
\usepackage{booktabs}
\usepackage{multirow}
\usepackage{tikz}
\usepackage{hyperref}
\hypersetup{
    colorlinks=true,
    linkcolor=purple,
    filecolor=magenta,      
    citecolor=blue
}
\usepackage{amsmath}
\usepackage{xcolor}
\usepackage{orcidlink}
\usepackage{graphicx,epsfig}
\usepackage{graphicx,subfigure}
\usepackage{commath}

\makeatletter
\long\def\dddddot#1{%
  {\mathop {#1}\limits ^{\vbox to-1.4\ex@ {\kern -\tw@ \ex@ \hbox {\normalfont .....}\vss }}}%
}
\long\def\multidots#1#2{%
  \count@=0
  {{\mathop {#2}\limits ^{\vbox to-1.4\ex@ {\kern -\tw@ \ex@ \hbox {\normalfont %
  \loop%
  \ifnum#1>\count@%
  .%
  \advance\count@ by1%
  \repeat%
  }\vss }}}}%
}
\makeatother

\begin{document}

\title{\bf Accelerating behavior from dynamical system analysis parameters}

\author{Rahul Bhagat\orcidlink{0009-0001-9783-9317}}
\email{rahulbhagat0994@gmail.com}
\affiliation{Department of Mathematics, Birla Institute of Technology and Science, Pilani, Hyderabad Campus, Jawahar Nagar, Kapra Mandal, Medchal District, Telangana 500078, India.}
\author{B. Mishra\orcidlink{0000-0001-5527-3565}}
\email{bivu@hyderabad.bits-pilani.ac.in}
\affiliation{Department of Mathematics, Birla Institute of Technology and Science, Pilani, Hyderabad Campus, Jawahar Nagar, Kapra Mandal, Medchal District, Telangana 500078, India.}

\begin{abstract}
We have performed the dynamical system analysis to obtain the critical point in which, the value of the geometric and dynamical parameters satisfy the late-time cosmic behavior of the Universe. At the outset, the modified Friedmann equations have been reformulated into a system of coupled differential equations to ensure that the minimal set of equations required for a second-order $f(Q)$ gravity. Then these equations are solved numerically to constrain the parameters with Markov Chain Monte Carlo (MCMC) techniques. Cosmic Chronometers (CC) and high-precision Pantheon$^+$ Type Ia Supernovae datasets are used to constrain the parameters. The evolution of key cosmological parameters indicates that the model exhibits quintessence-like behavior at present, with a tendency to converge towards the $\Lambda$CDM model at late-times. The dynamic system analysis provided the critical points that correspond to different phases of the Universe, which are analyzed in detail. The existence of a stable de Sitter attractor confirms the accelerating behavior of the model. 
\end{abstract}

\maketitle

\section{Introduction} 

Compelling evidence for the late-time accelerated expansion of the Universe first emerged from the High-Z Supernova Search Team\cite{Riess_1998_116} and the Supernova Cosmology Project\cite{Perlmutter_1998_517}. These discoveries were soon corroborated by a broad suite of independent observations, including the Cosmic Microwave Background (CMB)\cite{Giostri_2012_2012_027,Balkenhol_2023_108}, data from the Wilkinson Microwave Anisotropy Probe (WMAP)\cite{Larson_2011_192,Hinshaw_2013_208,Bennett_2003_148,Spergel_2003_148}, the Planck satellite\cite{Ade_2016_594a,Aghanim_2018_641}, large-scale structure data from the Sloan Digital Sky Survey (SDSS)\cite{Alam_2017_470,Eisenstein_2005_633}, and more recently, the Dark Energy Spectroscopic Instrument (DESI)\cite{Abbott_2018_480}. Collectively, these observations suggest that a mysterious component dark energy is driving the accelerated expansion of the Universe\cite{Abbott_2016_460}. In the standard cosmological framework, this phenomenon is attributed to the cosmological constant $\Lambda$, yielding the widely accepted $\Lambda$CDM model \cite{Riess_2019_876}, where dark energy has a constant equation of state $\omega_\Lambda = -1$.\\

While the $\Lambda$CDM model successfully explains a wide range of observations from the CMB to supernovae and baryon acoustic oscillations \cite{Raichoor_2020_500}, it is not without theoretical drawbacks. Two pressing concerns are the cosmic coincidence problem \cite{Cai_2005_2005_002}, which asks why the energy densities of matter and dark energy are comparable today despite their vastly different evolutionary paths, and the cosmological constant problem\cite{Weinberg_1989_61}, which highlights a striking discrepancy between the observed vacuum energy and predictions from quantum field theory \cite{Lobo_2009_173}. These unresolved issues have motivated extensive exploration of dynamical dark energy models. In these frameworks, the dark energy density evolves with time, as in scalar field theories like quintessence\cite{Carroll_1998_81,Bhagat_2023_42}, K-essence\cite{Picon_2001_63_103510}, or dark fluid models such as Chaplygin gas \cite{Odintsov_2018_98}. Such models allow for a time-varying equation of state parameter $\omega$, with quintessence typically constrained within $-1 \leq \omega \leq 1$, and phantom models \cite{Hrycyna_2007_76} extending into $\omega < -1$ regimes, enabling a more flexible interpretation of cosmic acceleration.\\

In parallel, another compelling line of inquiry involves modified gravity theories, which aim to explain cosmic acceleration without introducing unknown energy components. These approaches often extend or replace the Ricci scalar $R$ in Einstein's General Relativity (GR) with alternative geometric quantities. For example, teleparallel gravity formulates gravity in terms of the torsion scalar $T$, while symmetric teleparallel gravity also known as the symmetric teleparallel equivalent of GR (STEGR) employs the nonmetricity scalar $Q$\cite{Jimenez_2018_98,Cai_2016_79}. So, shifting the geometric focus from curvature and torsion to nonmetricity, described by the nonmetricity tensor $Q_{\lambda\mu\nu}$, can be defined as,

\begin{equation*}
Q_{\lambda\mu\nu} = \nabla_\lambda g_{\mu\nu} = \partial_\lambda g_{\mu\nu} - g_{\nu\sigma}\Gamma^\sigma_{\mu\lambda} - g_{\sigma\mu}\Gamma^\sigma_{\nu\lambda}.
\end{equation*}

It characterizes variations in the norm of vectors under parallel transport governed by a symmetric connection $\Gamma^\lambda_{\mu\nu}$. In STEGR, the absence of torsion and curvature shifts the interpretation of gravity to changes in lengths, offering a new geometrical perspective. The first significant extension of this framework is $f(Q)$ gravity, which generalizes the action to be a function of the nonmetricity scalar. This formulation provides a promising pathway to address the cosmic acceleration issue. Recent studies in $f(Q)$ gravity have significantly expanded its relevance both in cosmological and astrophysical contexts. Jimenez et al. \cite{Jimenez_2020_101} explored modified gravity theories based on the nonlinear extensions of the nonmetricity scalar $Q$ and proposed the  scenarios that can accommodate accelerating cosmologies relevant to both inflation and dark energy. Narawade et al. \cite{Narwade_2025_409} reconstructed viable $ f(Q) $ models based on novel forms of the deceleration parameter and demonstrated compatibility with observational values of the Hubble constant $ H_0 $ and matter density $ \Omega_{m0} $. Meanwhile, Heisenberg \cite{HEISENBERG20241} provided a foundational and pedagogical overview of metric-affine geometry, laying the mathematical groundwork for the geometric trinity of gravity and its extensions. Dimakis et al. \cite{Dimakis_2025_273} investigated power-law $ f(Q) $ models and scalar-nonmetricity theories, deriving exact solutions using a point-like Lagrangian in a non-coincident gauge. Vignolo et al. \cite{Vignolo_2024} approached the theory geometrically, introducing distribution-valued tensors to derive junction conditions for smoothly connecting spacetime regions in $ f(Q) $ gravity. Furthermore, Paliathanasis \cite{Paliathanasis_2024_46} examined tilted perfect fluids in the Kantowski-Sachs metric, revealing that the evolution of the tilt is tightly constrained by the non-linear structure of the field equation.\\

In the astrophysical domain, Alwan et al.\cite{Alwan_2024} studied the structure of neutron stars within covariant $ f(Q) $ gravity and examined the mass-radius relations for various functional forms of $ f(Q) $. Jensko \cite{Jensko_2025} contributed to the cosmological application of symmetric teleparallel gravity by constructing FLRW models using coincident gauge coordinates aligned with cosmological Killing vectors, identifying both flat and curved spatial solutions. Chen \cite{Chen_2025} employed the topological RVB formalism to compute Hawking temperatures in $ f(Q) $ black hole solutions, revealing additional contributions to black hole thermodynamics. Rastgoo et al. \cite{Rastgoo_2024_84} showed that traversable wormhole solutions can be sustained within $ f(Q) $ gravity without the need for exotic matter. Agrawal et al.\cite{Agrawal_2023_83} extended the theory to a bouncing cosmological scenario inspired by loop quantum cosmology, demonstrating a viable matter-bounce scenario within extended symmetric teleparallel gravity. Collectively, these contributions underscore the potential of $ f(Q) $ gravity as a compelling alternative to GR and capable of addressing both gravitational phenomena and late-time cosmic acceleration \cite{Capozziello_2022_832,Najera_2023_524,Ambrosio_2022_105,BHAGAT2025101913}.\\

In this paper, we aim to frame the cosmological model of the Universe that supports the late-time accelerating behavior of the Universe using two key observational datasets, Cosmic Chronometers (CC) \cite{Moresco_2015_450} and Pantheon$^+$ sample \cite{Scolnic_2022}. The paper is organized as: Sec. \ref{Sec:2}, the modified field equations of $ f(Q) $ gravity has been presented. In Sec. \ref{Sec:3}, the observational analysis and process are described in detail and the results obtained using the numerical approach are listed. In Sec. \ref{Sec:4}, we perform the dynamical system analysis to obtain the critical points. The results and conclusions are given in Sec. \ref{Sec:5}.

\section{Field Equations of $f(Q)$ gravity}\label{Sec:2}

The action of $f(Q)$ gravity \cite{Jimenez_2018_98} given as,
\begin{equation}\label{Eq:1}
    S = \frac{1}{2} \int d^4x \sqrt{-g} f(Q) + S_m,
\end{equation}
where $ 8\pi G/c^4 = 1 $ and $ S_m $ represents the matter action. Varying the action with respect to the metric yields the following field equations,

\begin{equation}\label{Eq:2}
      \frac{2}{\sqrt{-g}}\nabla_\alpha \left( \sqrt{-g} f_Q P^\alpha_{~~\mu\nu} \right) - \frac{1}{2} f g_{\mu\nu} + f_Q \left( P_{\mu\alpha\beta} Q_{\nu}^{~~\alpha\beta} - 2  Q^{\alpha\beta}_{~~~\mu}P_{\alpha\beta\nu} \right) = T_{\mu\nu},
\end{equation}

where $ f_Q = \frac{\partial f}{\partial Q} $. Varying the action with respect to the connection yields,

\begin{equation}\label{Eq:3}
    \nabla_\mu \nabla_\nu \left( \sqrt{-g} f_Q P^{\mu\nu}_{~~~\alpha}\right) = 0,
\end{equation}

which serve as dynamical equations for the connection. The nonmetricity conjugate tensor $ P^\alpha_{\mu\nu} $ is defined as,

\begin{equation}\label{Eq:4}
    P^{\alpha}_{~~\mu\nu} = -\frac{1}{4} Q^{\alpha}_{\mu\nu} + \frac{1}{2} Q_{{(\mu}}{ \alpha_{\nu)}} + \frac{1}{4} (Q^\alpha - \tilde{Q}^\alpha) g_{\mu\nu} - \frac{1}{4} \delta^{\alpha}_{~{(\mu}}{ Q_{\nu)}},
\end{equation}

where symmetrization of the indices is denoted as $ A_{(\mu\nu)} = \frac{1}{2} (A_{\mu\nu} + A_{\nu\mu}) $. The independent traces of the nonmetricity tensor are given as,

\begin{equation}\label{Eq:5}
    Q_\alpha = g^{\sigma\lambda} Q_{\alpha\sigma\lambda}, \quad \tilde{Q}_\alpha = g^{\sigma\lambda} Q_{\sigma\alpha\lambda}.
\end{equation}
Now, the nonmetricity scalar $Q$ can be expressed as,
\begin{equation}\label{Eq:6}
Q = -Q_{\alpha\mu\nu} P^{\alpha\mu\nu}.
\end{equation}

We consider homogeneous and isotropic flat FLRW space time as,

\begin{equation}\label{Eq:7}
    ds^2 = -dt^2 + a^2(t) \left[dx^2 + dy^2 + dz^2\right], 
\end{equation}

where $ a(t) $ denotes the scale factor. The chosen gauge in this line element, known as the coincident gauge, ensures that the metric is the sole fundamental variable. However, selecting a different gauge introduces a nontrivial contribution to the field equations due to the connection properties \cite{Jimenez_2018_98,HEISENBERG20241}. For the metric \eqref{Eq:7}, the nonmetricity scalar $Q$ can be obtained as,
\begin{equation}\label{Eq:8}
Q = 6H^2,
\end{equation}
where $H=\frac{\dot{a}}{a}$ is the Hubble parameter. The matter content is considered to be that of a perfect fluid, characterized by the energy momentum tensor,

\begin{equation}\label{Eq:9}
T_{\mu\nu} = (\rho + p)u_\mu u_\nu + p g_{\mu\nu},
\end{equation}

where $\rho$ and $p$ respectively, represent the energy density and pressure, $ u_\mu $ is the four-velocity vector of the fluid. Now, the field equations of $f(Q)$ gravity \eqref{Eq:2} for the space time \eqref{Eq:7}, can be written as,
\begin{eqnarray}\label{Eq:10}
6H^2~f_Q - \frac{1}{2}~f &=& \rho,  \\
(12H^2f_{QQ}+f_{Q})\dot H &=& -\frac{1}{2}(\rho + p).\label{Eq:11}
\end{eqnarray}

These equations govern the dynamics of the system and will serve as the basis for further study. If we consider the transformation, $f(Q)= Q + \Phi(Q)$, Eq. \eqref{Eq:10} and Eq. \eqref{Eq:11} reduce to, 

\begin{equation}\label{Eq:12}
     3H^2 = \rho + \frac{1}{2}\Phi- Q~\Phi_Q
\end{equation}
\begin{equation}\label{Eq:13}
     -3H^2 - 2\dot{H} = p + Q~\Phi_Q - \frac{1}{2}\Phi+2\dot{H}~(2Q~\Phi_{QQ}+\Phi_Q)
\end{equation}
To frame the cosmological model, we need to assume some functional form of $f(Q)$, and hence we adopted the well-motivated logarithmic form. 
\subsection{Logarithmic f(Q) Model}
The functional form of $f(Q)$ can drive the accelerated expansion of the Universe, offering a geometrical alternative to dark energy. Notably, General Relativity (GR) with a cosmological constant is recovered as a particular case when $f(Q) = Q + 2\Lambda$. Introducing a general $f(Q)$ modifies the Einstein-like field equations by adding terms that involves the derivatives of $f(Q)$, thereby enriching the cosmological dynamics. Among the various models explored in the literature, the power-exponential form proposed in \cite{Anagnostopoulos_2021_822,SULTANA2025100422,B_hmer_2023} demonstrates a range of interesting behaviors that can reproduce late-time cosmic acceleration and fit observational data effectively. Similarly, the power-law model examined by Paliathanasis \cite{PALIATHANASIS2025101993} leads to field equations that can be expressed in terms of an integrable Hamiltonian system, and under specific parameter constraints, it admits an analytical expression for the dynamical dark energy equation of state (EoS) parameter. Here, we consider a logarithmic-type function introduced by N\'ajera et al. \cite{Najera_2023_524} as,

$$
\Phi(Q) = \alpha~\frac{Q^{\beta+1}}{Q_0^{\beta}}\log\left(\frac{Q}{Q_0}\right),
$$

where $\alpha$ and $\beta$ are free parameters of the model. In the original formulation by N\'ajera et al.\cite{Najera_2023_524}, the parameters carry physical dimensions. Here, we rescale and modify the model to use dimensionless parameters, facilitating a more convenient numerical implementation and enabling a direct comparison with observational data using dimensionless, model-independent variables. The logarithmic dependence on $Q$ allows deviation from the linear GR term at low curvatures while still permitting significant modifications at higher curvature regimes, making it suitable for describing both early and late-time cosmic dynamics within a unified framework. The exponent $\beta$ controls the rate of deviation from GR, while $\alpha$ sets the overall strength of the modification. The inclusion of the logarithmic term can also lead to scale-dependent effects that are absent in simple polynomial models, thereby offering greater flexibility in fitting a broad range of cosmological observations. The dimensionless parameters approach simplifies the dynamical equations and makes parameter space exploration through MCMC analysis more efficient. Now, to frame the cosmological model using numerical methods, we need to assume some appropriate sets of dimensionless and model-independent variables,

\begin{equation}\label{Eq:14}
h=\frac{H}{H_0},\ \ \, x=\frac{\rho_m}{3H^2},\ \ \ y=\frac{\rho_r}{3H^2}.
\end{equation}

The Universe is filled with dust and radiation fluids, hence
\begin{equation}\label{Eq:15}
\rho = \rho_m + \rho_r, \ \ \ \ \ p_r=\frac{1}{3}\rho_r,\ \ \ \ p_m=0.
\end{equation}
The density parameters for matter, radiation and dark energy are respectively,
\begin{equation}\label{Eq:16}
\Omega_m=\frac{\rho_m}{3 H^2},~~~\Omega_r=\frac{\rho_r}{3 H^2},~~~\Omega_{f}=\frac{ \rho_{eff}}{3 H^2}
\end{equation}

Now, using Eq. \eqref{Eq:14} - Eq.\eqref{Eq:16}, Eq.(\ref{Eq:12}) and Eq.(\ref{Eq:13})  can be respectively rewritten as,
\begin{equation}\label{Eq:17}
1 = \Omega_{m} + \Omega_{r} +\alpha  \left(-h^{2 \beta }\right) \left((2 \beta +1) \log \left(h^2\right)+2\right)
\end{equation}
\begin{equation}\label{Eq:18}
    -\frac{2\dot{H}}{3H^2}=\frac{\frac{4}{3} \left(2 \alpha  h^{2 \beta }+\alpha  h^{2 \beta } \log \left(h^2\right)+2 \alpha  \beta  h^{2 \beta } \log \left(h^2\right)-x+1\right)+x}{\alpha  h^{2 \beta } \left((\beta +1) \log \left(h^2\right)+1\right)+2 \alpha  h^{2 \beta } \left(2 \beta +(\beta +1) \beta  \log \left(h^2\right)+1\right)+1}
\end{equation}

Also, we can obtain the deceleration parameter and total EoS parameter as,

\begin{equation}\label{Eq:19a}
  q=  -\frac{2 \alpha  h^{2 \beta } \left(4 \beta +(\beta -1) (2 \beta +1) \log \left(h^2\right)-1\right)+x-2}{2 \alpha  h^{2 \beta } \left(4 \beta +(\beta +1) (2 \beta +1) \log \left(h^2\right)+3\right)+2},
\end{equation}

\begin{equation}\label{Eq:19b}
    w=-\frac{\alpha  h^{2 \beta } \left(12 \beta +\left(6 \beta ^2+\beta -1\right) \log \left(h^2\right)+1\right)+x-1}{3 \alpha  h^{2 \beta } \left(4 \beta +(\beta +1) (2 \beta +1) \log \left(h^2\right)+3\right)+3}
\end{equation}\\

Taking derivatives of $x$ and $h$ with respect to the redshift z and using Eq. (\ref{Eq:17}) and Eq. (\ref{Eq:18}), we can obtain two coupled differential equations.

    \begin{equation}\label{Eq:20}
    \frac{dx}{dz}=\frac{x(z) \left(\alpha  h(z)^{2 \beta } \left(12 \beta +\left(6 \beta ^2+\beta -1\right) \log \left(h(z)^2\right)+1\right)+x(z)-1\right)}{\alpha  (z+1) h(z)^{2 \beta } \left(4 \beta +(\beta +1) (2 \beta +1) \log \left(h(z)^2\right)+3\right)+z+1},
\end{equation}
\begin{equation}\label{Eq:21}
    \frac{dh}{dz}=\frac{h(z) \left(4 \alpha  h(z)^{2 \beta } \left((2 \beta +1) \log \left(h(z)^2\right)+2\right)-x(z)+4\right)}{2 (z+1) \left(\alpha  h(z)^{2 \beta } \left(4 \beta +(\beta +1) (2 \beta +1) \log \left(h(z)^2\right)+3\right)+1\right)}.
\end{equation}

Here, our objective is to determine the initial condition for the parameter $x$ and the model parameters $\alpha$ and $\beta$. In the next section, we shall use the cosmological datasets to constrain these parameters.

\section{Observational Analysis and the Results}\label{Sec:3}
We shall numerically solve the system of coupled differential equations outlined in Eqs. (\ref{Eq:20}) and (\ref{Eq:21}), using initial conditions $x_0=\Omega_{m_0}$ and $h_0 = 1$. Within the framework of the considered $f(Q)$ gravity model, this setup introduces four independent free parameters, the present day Hubble constant $H_0$, the matter density parameter $\Omega_{m0}$, and the model parameters $\alpha$ and $\beta$. We constrain these parameters via the MCMC technique, embedded within a Bayesian inference framework \cite{Foreman-Mackey_2013_125}. Bayesian analysis requires the specification of prior ranges to capture potential parameter space adequately [Table \ref{tableA1} {\bf(Last Row)}]. The parameter constraints are informed by two complementary datasets: the Hubble parameter measurements from CC and the luminosity distance data from Type Ia Supernovae (SNIa). For CC, we adopt the dataset from Ref. \cite{Moresco_2022_25}, while for SNIa, we utilize the comprehensive Pantheon$^+$ compilation \cite{Brout_2022_938}. The CC dataset consists of 32 measurements of $H(z)$ in the redshift range $0.07 \leq z \leq 1.965$, while the Pantheon$^+$ sample encompasses 1,701 light curves corresponding to 1,550 distinct Type Ia supernovae distributed across  $0.00122 \leq z \leq 2.2613$ \cite{Scolnic_2022.03863,Brout_2022_938}. The observed magnitude-redshift relation for SNIa is modeled through,
\begin{equation}\label{Eq:22}
\mu(z) = 5 \log_{10} \left(\frac{d_L(z)}{\text{Mpc}}\right) + \mu_0,
\end{equation}
where $\mu_0$ is the nuisance parameter,

  \begin{equation}
      \mu_0 = 25 + 5 \log_{10} \left( \frac{c/H_0}{\text{Mpc}} \right),
  \end{equation}
  
and $d_L(z)$ is the luminosity distance. In spatially flat FLRW cosmology ($\Omega_k = 0$), the luminosity distance becomes,
\begin{equation}\label{Eq:23}
d_L(z) = c (1 + z) \int_0^z \frac{dz'}{H(z')}, 
\end{equation}
where $c$ denotes the speed of light in $\text{km/s}$. To quantify the agreement between model predictions and data, we employ chi-squared statistics. For the CC dataset, the chi-squared is defined as,
\begin{equation}\label{Eq:24}
\chi^2_\text{CC} = \sum_{i=1}^{32} \left( \frac{H_\text{obs}(z_i) - H_\text{th}(z_i, H_0, \Omega_{m0}, \alpha, \beta)}{\sigma(z_i)} \right)^2, 
\end{equation}
where $H_\text{obs}$ and $H_\text{th}$ respectively be the observed and theoretical Hubble rates, and $\sigma(z_i)$ represents the associated uncertainties. For the Pantheon$^+$ dataset, the corresponding chi-squared statistic is,
\begin{equation}\label{Eq:25}
\chi^2_\text{SNIa} = \Delta \mathbf{D}^T \mathbf{C}_\text{SNIa}^{-1} \Delta \mathbf{D}, 
\end{equation}
where $\Delta \mathbf{D}$ is the residual vector of the distance moduli and can be expressed as,
\begin{equation}\label{Eq:26}
\Delta \mathbf{D} = \mu(z_i) - \mu_\text{th}(z_i, H_0, \Omega_{m0}, \alpha, \beta), 
\end{equation}
and $\mathbf{C}_\text{SNIa}$ is the full covariance matrix. For MCMC sampling, we use the Python package \texttt{emcee}, which implements the affine invariant ensemble sampler Ref. \cite{Foreman-Mackey_2013_125}.

\subsection{MCMC Result}
After performing the MCMC analysis, we constrain the model parameters using the CC dataset, the Pantheon$^+$ dataset, and the combined CC+Pantheon$^+$ supernova dataset. The resulting confidence contours for these parameters are illustrated in Fig. \ref{Fig1}. The best-fit values along with their uncertainties are summarized in Table \ref{tableA1}. The contour plots provide visual insight into how well the parameters are constrained by each dataset and reveal correlations between them. For the CC dataset alone, the contours are relatively broad, indicating weaker constraints due to the limited number of observational points, leading to a larger allowed parameter space. In contrast, the Pantheon$^+$ dataset, with 1701 high-quality supernova data points, offers much tighter constraints with smaller and more elongated contours, indicating strong parameter correlations and improved precision. When the two datasets are combined, the contours shrink and become more refined, especially for parameters that are sensitive across different redshift ranges. This clearly shows the advantage of using complementary datasets. while CC data helps to constrain high-redshift behavior, Pantheon$^+$ provides precise constraints at low redshifts. The synergy between them results in tighter bounds, reduced degeneracies, and a more robust determination of cosmological parameters.

To further validate the consistency of the model with observational data, we present error bar plots for the Hubble parameter $H(z)$ corresponding to the CC data and the distance modulus $\mu(z)$ for the Pantheon$^+$ dataset in Fig. \ref{Fig2}. These plots visually demonstrate the alignment between the theoretical predictions of the model and the actual observational data. The model curves lie well within the error margins of the data, supporting the statistical results obtained from the MCMC analysis and confirming the viability of the model in describing the expansion history of the Universe.

\begin{widetext}
    \begin{figure}[H]
    \centering
    \includegraphics[width=17cm,height=15cm]{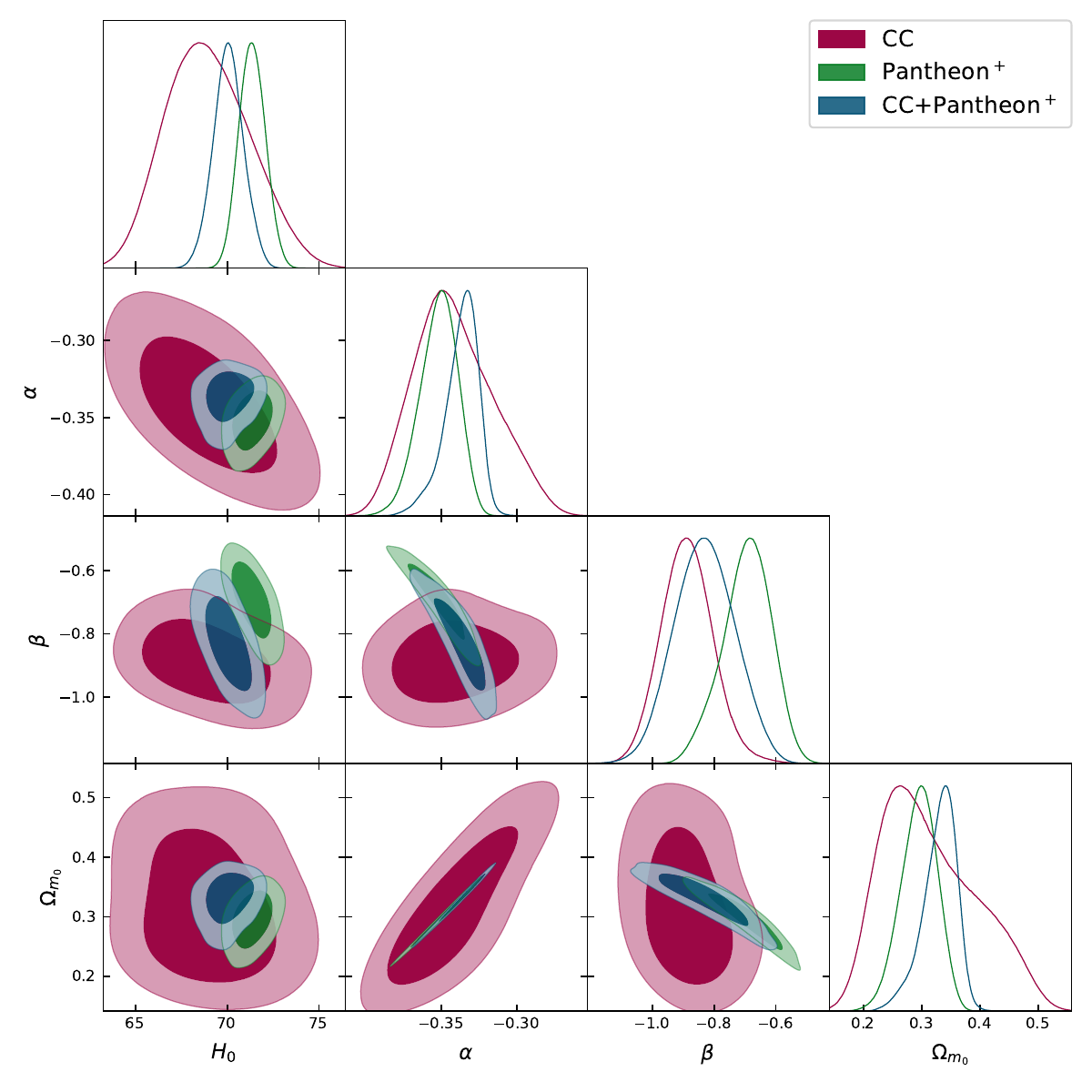}

    \caption{ The contour plots for 32 points of CC sample, 1701 light curves from Pantheon$^+$ dataset and combine CC+Pantheon$^+$ dataset upto 2$-\sigma$ errors for the parameters $H_0$, $\Omega_{m_0}$, $\alpha$ and $\beta$.
} 
    \label{Fig1}
\end{figure} 

\begin{table*}[htb]
\renewcommand\arraystretch{1.5}
\centering 

\begin{tabular}{|c|c|c|c|c|c|c|c|c|c|c|} 
\hline 
~~~Data sets~~~& ~~~$H_0$  ~~~& ~~~~$\alpha$~~~~ & ~~~$\beta$~~~& ~~~~$\Omega_{m_0}$~~~~&  \parbox[c][1.5cm]{2cm}{Age of the Universe (Gyr)} &~~~$q_0$~~~&~~~$z_{tr}$~~~&~~~$w_0$~~~&~~~$r_0$~~~&~~~$s_0$~~~
     \\ [0.5ex] 
\hline\hline
CC  & $68.9^{+2.1}_{-2.5}$ &  $-0.342^{+0.023}_{-0.030}$ & $-0.888^{+0.076}_{-0.085}$&$0.314^{+0.060}_{-0.10}$& $13.913$& $-0.600$&$0.681$&$-0.733$ &$1.040$& $-0.012$ \\
\hline
 Pantheon$^+$ & $71.35\pm0.71$ &  $-0.353^{+0.014}_{-0.011}$ & $-0.712^{+0.088}_{-0.069 }$&$0.295^{+0.034}_{-0.029}$& $13.671$& $-0.535$&$0.696$&$-0.690$ &$0.960$& $0.0127$ \\
\hline
 CC+Pantheon$^+$ & $70.06\pm0.82$ &  $-0.337^{+0.013}_{-0.0079}$ & $-0.830\pm0.096$&$0.326^{+0.038}_{-0.016}$& $13.527$& $-0.558$&$0.651$&$-0.705$ &$1.017$& $-0.005$ \\
\hline
Priors & $(60,80)$ &  $(-2,2)$ & $(-2,2)$& $(0.1,0.7)$&$-$&$-$&$-$&$-$&$-$&$-$\\
 \hline 
\end{tabular}
\caption{Constrained parameter values, and present value of cosmological parameters for CC, Pantheon$^+$ and CC+Pantheon$^+$ data sets.} 
\label{tableA1}
\end{table*}

\end{widetext}

\begin{widetext}
    
\begin{figure}[H]
\centering
\includegraphics[width=12.2cm,height=6cm]{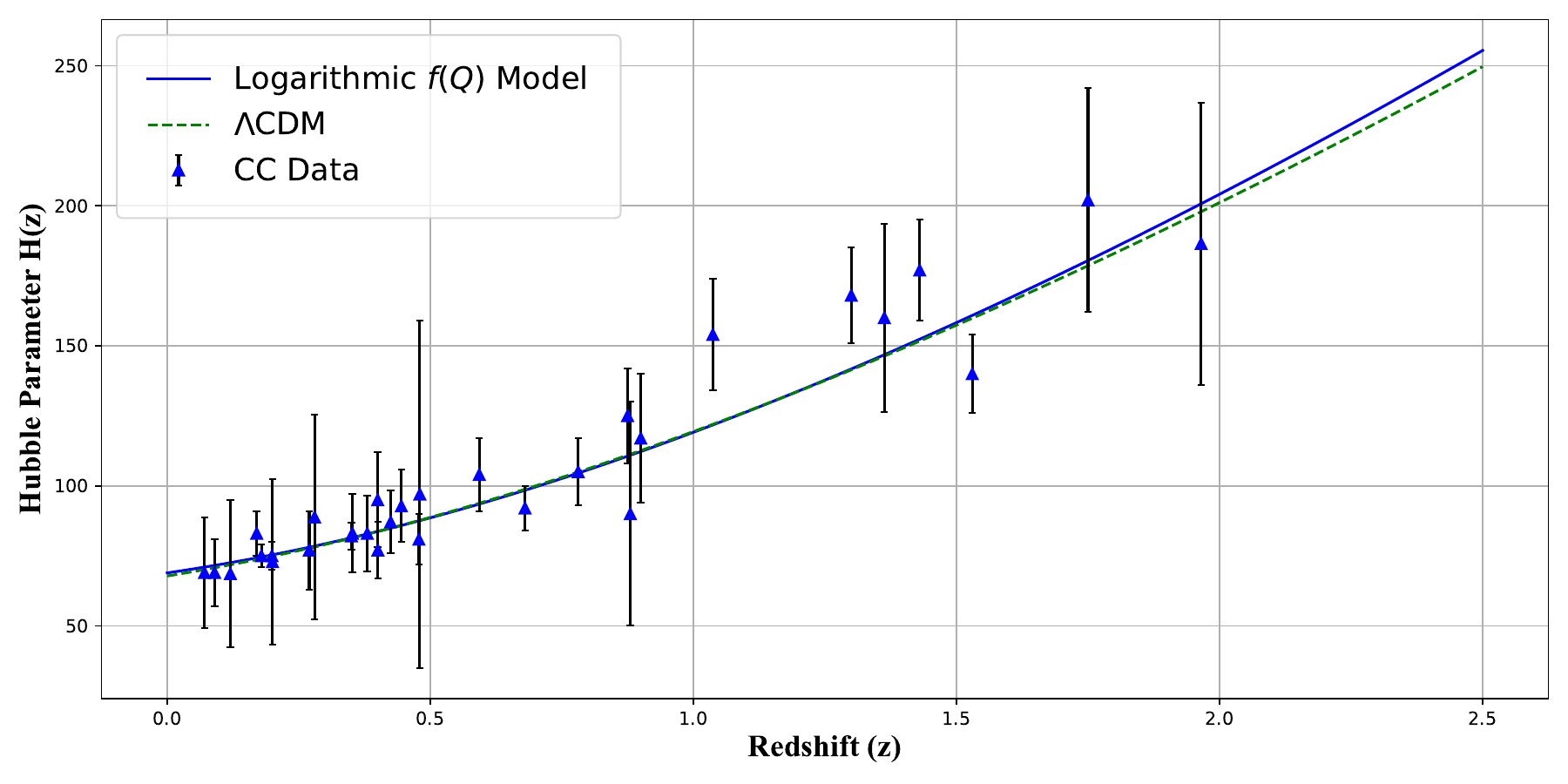}
\includegraphics[width=14.3cm,height=7cm]{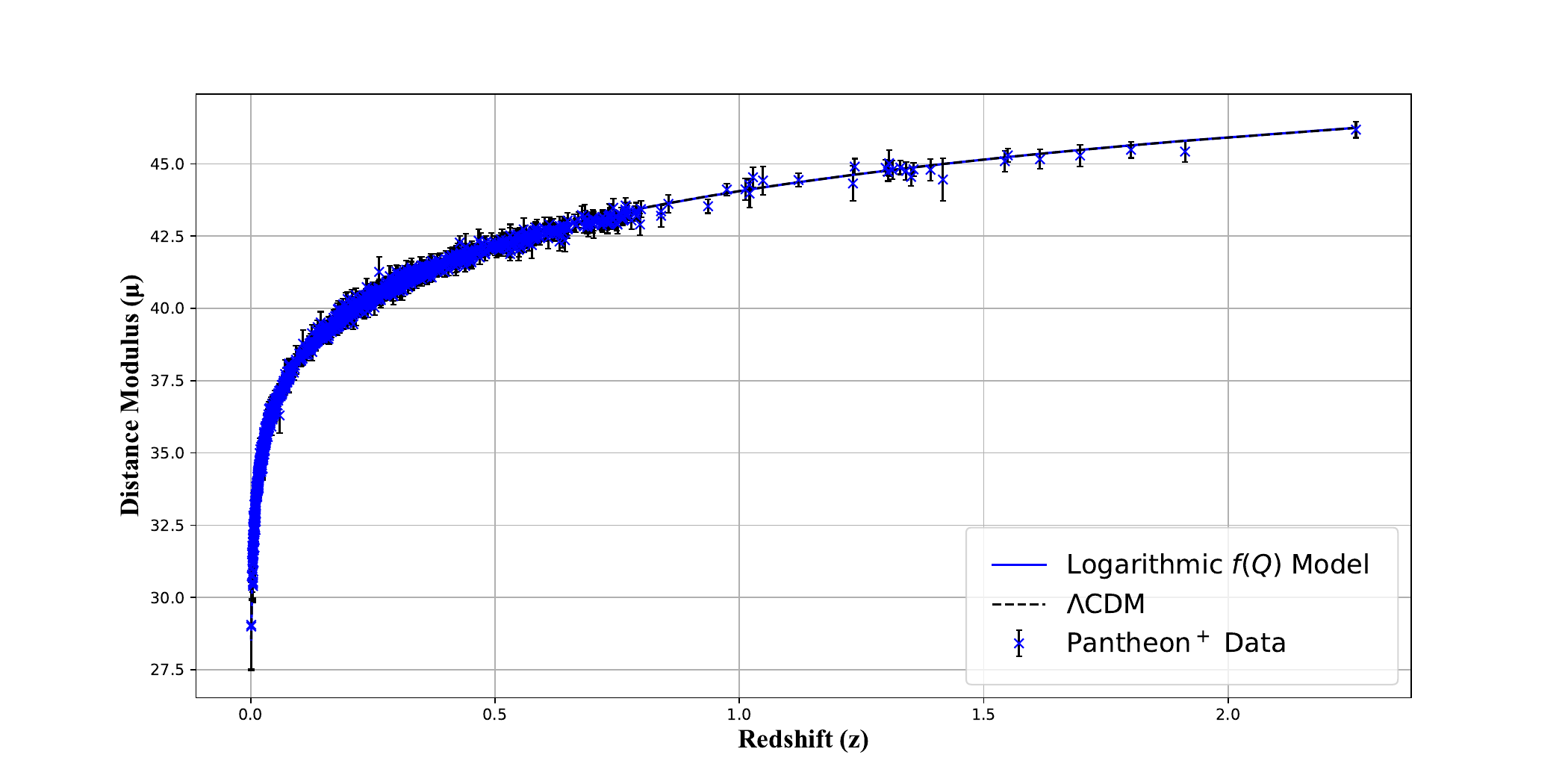}
\caption{Evolution of Hubble parameter \textbf{(Top Panel)} and distance modulus parameter \textbf{(Bottom Panel)} for the model and $\Lambda$CDM.}
\label{Fig2}
\end{figure}

\end{widetext}

\subsection{Cosmological Parameters}

\begin{widetext}

 \begin{figure}[H]
    \centering
   
        \includegraphics[width=12cm]{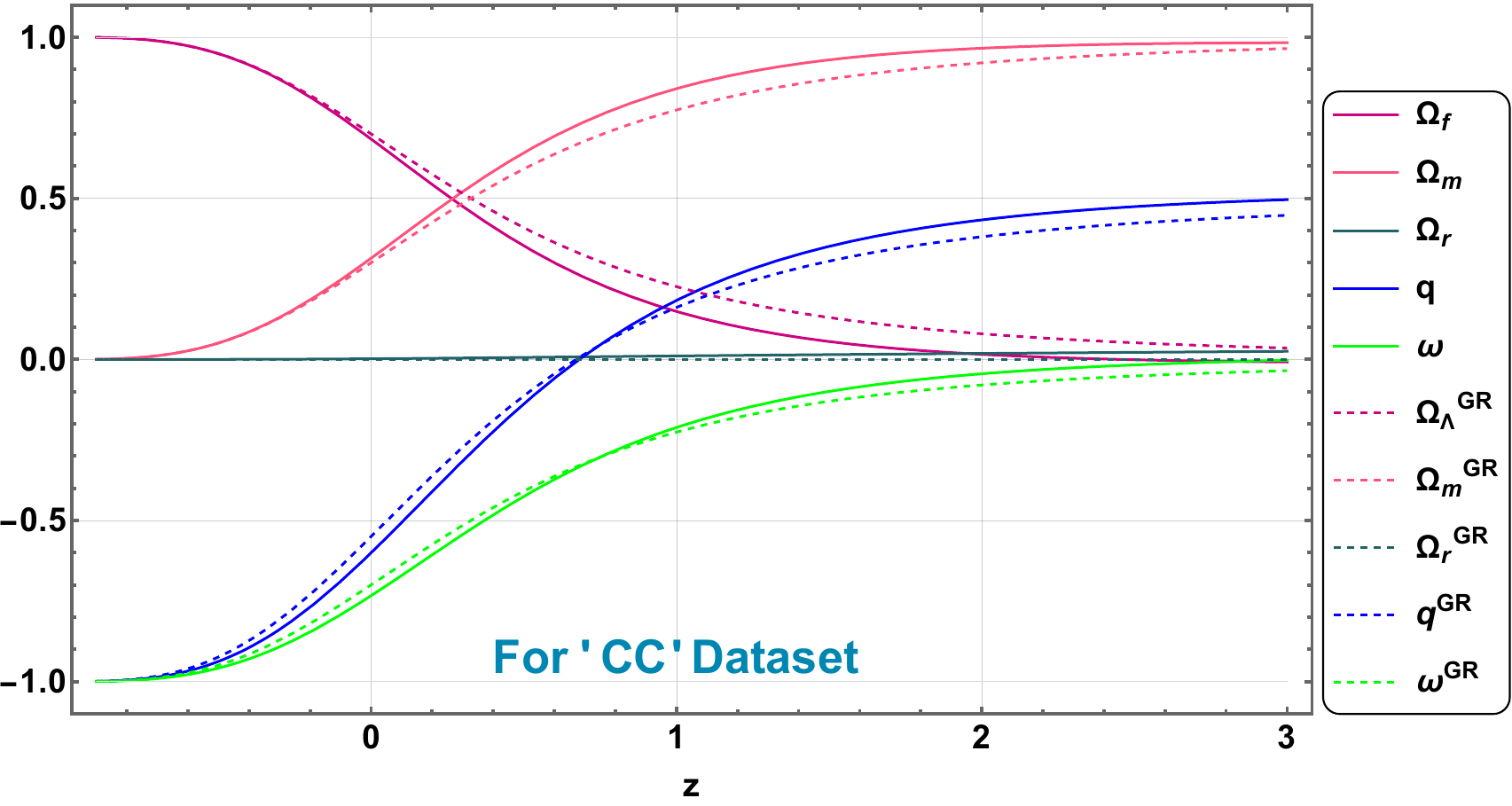}

        \includegraphics[width=12cm]{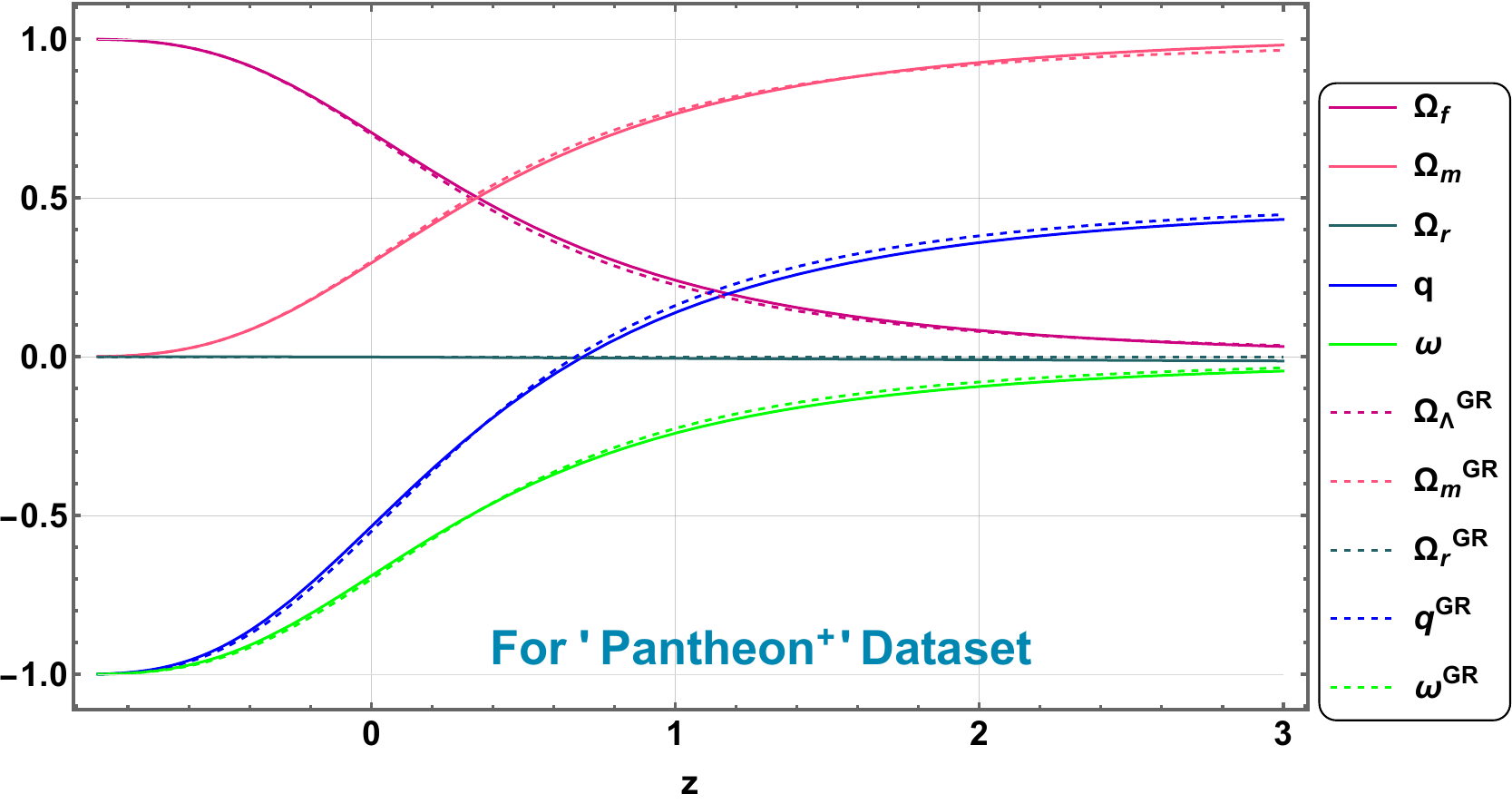}

      \includegraphics[width=12cm]{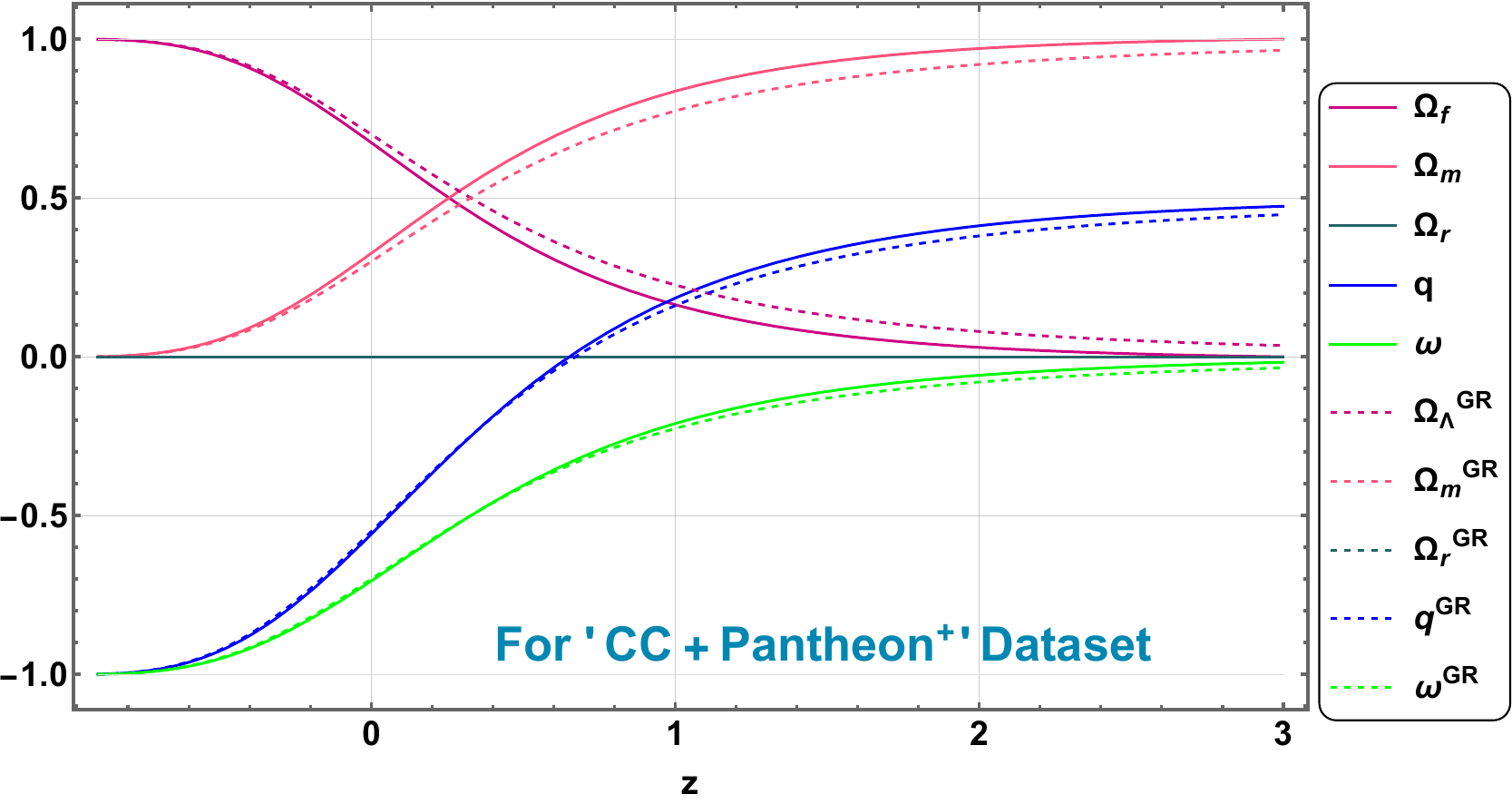}  

    \caption{Evolution of Cosmological parameters for observational Dataset (Upper Panel for CC, Middle Panel for Pantheon$^+$ and Lower Panel for CC+Pantheon$^+$ Dataset) and dotted lines show it comparison to GR with a positive cosmological constant.} 
    \label{Fig3}
\end{figure}
\end{widetext}

After obtaining the best-fit values of the model parameters, we have examined the cosmological implications by analyzing the evolutionary behavior of density parameters, equation of state (EoS) parameter and deceleration parameter. Fig. \ref{Fig3} presents this analysis in three panels. The upper panel shows the evolution of these parameters using the CC dataset, while the middle and lower panels correspond to the Pantheon$^+$ dataset and the combined CC+Pantheon$^+$ dataset, respectively. A comparison with the standard $\Lambda$CDM model is also included for reference. We observe that the evolution of the density parameters $\Omega_m$, $\Omega_r$, and $\Omega_f$ in the Pantheon$^+$ case closely follows the $\Lambda$CDM predictions. In contrast, the CC dataset comprising only 32 data points exhibits slightly broader uncertainty bands and mild deviations at early times. This is expected, as fewer data points result in weaker constraints on the model parameters. On the other hand, the Pantheon$^+$ dataset, which includes 1701 supernova data points, provides significantly tighter constraints. Further, we analyzed the combined CC + Pantheon$^+$ dataset, the effect of using the combined dataset is significantly improves the constraints on cosmological parameters by incorporating the strengths of both datasets. We observed that, in the combined datastes, the constraints are noticeably tighter than those obtained from CC dataset alone. As a result, the evolution of these parameters aligns more closely with the $\Lambda$CDM in late time, reducing uncertainties and enhancing the reliability of the model across the entire redshift range.  At the current epoch, the contribution of dark energy is approximately $ \Omega_f \approx 0.7 $, while the matter component accounts for about $ \Omega_m \approx 0.3 $, and radiation contributes a negligible amount for ($ \Omega_r $).

In the case of CC dataset, the present value of deceleration parameter, $q_0 = -0.600$, with a transition redshift $z_t = 0.681$. For the Pantheon$^+$ dataset, the values are $q_0 = -0.535$ and the transition noted at $z_t = 0.696$. Whereas for the combined datasets, we obtained $q_0 = -0.558$ and $z_t = 0.651$. These results consistently support the late-time acceleration of the Universe. The evolution of the EoS parameter $\omega(z)$, which characterizes the pressure-to-density ratio of the cosmic fluid further confirms this. At present, we find $\omega_0 = -0.733$ for the CC dataset, $\omega_0 = -0.690$ for the Pantheon$^+$ dataset, and $\omega_0 = -0.705$ for the combined dataset. These values approach the cosmological constant behavior ($\omega = -1$), suggesting a transition towards a de Sitter phase in the late Universe.

The state-finder pair serves as an important diagnostic tool to characterize and differentiate cosmological models. These parameters are effective in the probe of the nature of dark energy and modified gravity theories. The state finder pair can be obtained as (Ref. \cite{Sahni_2003_77})

\begin{equation}\label{Eq:27}
r = \frac{\dddot{a}}{aH^3}
\end{equation}
\begin{equation}
s = \frac{r - 1}{3(q - \frac{1}{2})} \label{Eq:28}    
\end{equation}  

\begin{figure}[H]
\centering
\includegraphics[width=9.9cm]{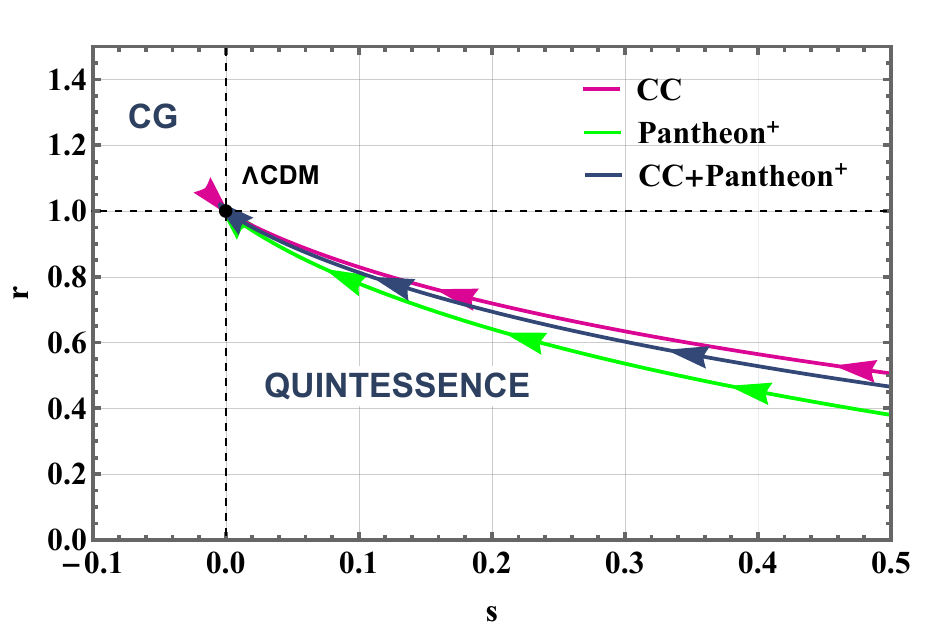}
\caption{Behavior of state finder parameter.} 
\label{Fig4}
\end{figure}

In $\Lambda$CDM framework, the state finder pair value prescribed to be $ r = 1 $ and $ s = 0 $ \cite{Alam_2003}, which represents the reference point in the $(r,s)$ plane. Any deviation from these values signals a departure from the standard cosmological model, offering a powerful means to distinguish alternative scenarios such as quintessence, phantom, Chaplygin gas and so on.
Fig. \ref{Fig4} represents the evolutionary behavior of state finder pair. 
The present values are found to be $ r_0 = 1.040 $ and $ s_0 = -0.012 $ for the CC dataset, $ r_0 = 0.960 $ and $ s_0 = 0.0127 $ for the Pantheon$^+$ sample, and $ r_0 = 1.017 $ and $ s_0 = -0.005 $ for the CC+Pantheon$^+$ sample . Over time, the trajectories of these parameters show convergence toward the $\Lambda$CDM point. From this, we conclude that the model shows quintessence-like behavior. 

To estimate the age of the Universe, we refer to the age redshift relation,

\begin{equation}\label{Eq:29}
t_U(z) = \int_{z}^{1089} \frac{dz'}{(1 + z') H(z')} \, 
\end{equation}  

\begin{figure}[H]
    \centering
    \includegraphics[width=9.5cm]{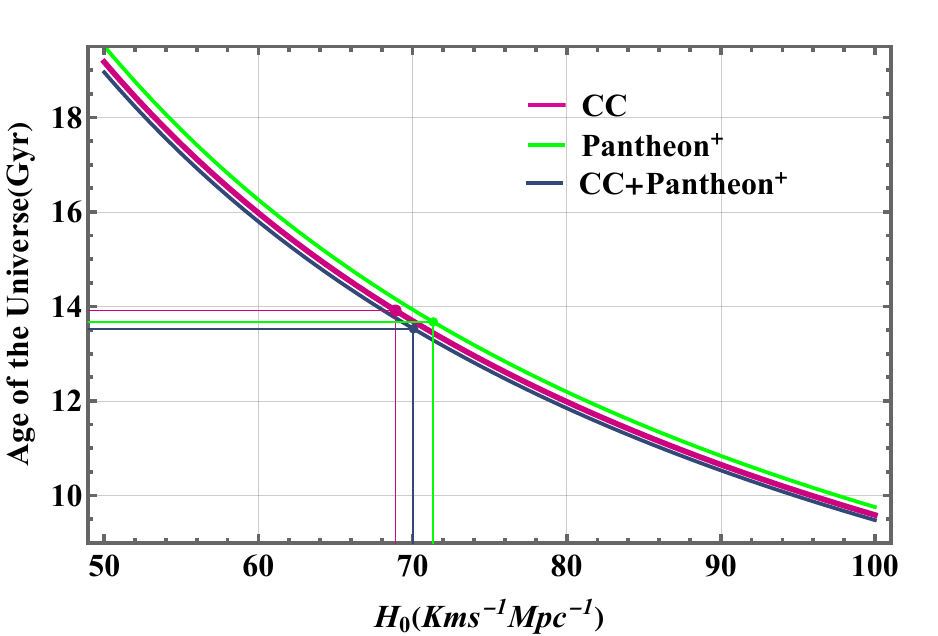}

    \caption{Plot of the Age of the Universe in Gyr.} 
    \label{Fig5}
\end{figure}

This integral quantifies the time elapsed from a given redshift $z$ to the surface of the last scattering ($z = 1089$). The present age of the Universe $ t_0 = t_U(0) $ depends inversely on the value of the Hubble constant $H_0$. In this model, we obtain the present age of the Universe approximately 13.913 Gyr (CC dataset), 13.671 Gyr (Pantheon$^+$ dataset) and 13.527 Gyr (CC+Pantheon$^+$ dataset). According to Ref. \cite{Age_Hubble} the estimated age is $ t_0 \approx 14.46 \pm$ 0.8 \text{Gyr}. Subsequently, Planck satellite datasets revealed the age as,  $ t_0 \approx 13.78 \pm 0.02 \, \text{Gyr} $ \cite{Plank_2020}. The stellar evolution models have revised as, the star age to around $ t_0 \approx 13.7 \text{Gyr}$ \cite{Creevey_2015}. Some more suggestions are : $t_0 \approx 13.7 \text{Gyr}$ \cite{Tang_2021}, $ t_0 \approx 13.8\pm 4 \text{Gyr}$ \cite{Cowan_2002}. The estimated value obtained through different datasets well within the prescribed values of the cosmological observations. The evolutionary behavior has been shown in Fig. \ref{Fig5}.

\section{Phase Space Analysis: Fixed Points and Late-Time Behavior}\label{Sec:4}

In this section, we shall study the stability of the model through dynamical system analysis. This will enable us to  reformulate the cosmological equations into an autonomous system of differential equations. It will facilitates to identify and classify the critical points. In some of the earlier works, it has been shown that in the context of symmetric connections in a spatially flat FLRW background, the de Sitter Universe consistently emerges as a late time attractor \cite{Paliathanasis_2023_41,Bhagat_ASPdyna2024}. Complementarily, B$\ddot{o}$hmer et al.\cite{Bohmer_2016_book_dyna,B_hmer_2022} provided a systematic introduction to dynamical systems methods emphasizing their utility in probing the qualitative behavior of complex cosmological models. These studies underscore the relevance of phase space analysis in understanding the evolutionary dynamics of modified gravity theories.  For this, we introduce four dimensionless variables, which are central to the formulation of the dynamical system. 

\begin{equation}\label{Eq.32}
x=\frac{\rho_m}{3H^2},\ \ \ y=\frac{\rho_r}{3H^2},\ \ \,\eta=\frac{\Phi}{Q},\ \ \,\delta=-2\Phi_Q.\ \ \ 
\end{equation}

Now, using dimensionless variables recasting the generalized Friedmann equations \eqref{Eq:12} and \eqref{Eq:13} in terms of these variables can be rewritten as,
\begin{equation}\label{Eq:17}
1 = x + y +\eta+\delta.
\end{equation}
Here, we eliminate the fourth variable $\delta$ to make our system in the form of three independent variables. We are following the equation from the second Friedmann equation,
\begin{equation}\label{Eq:18}
    -\frac{2\dot{H}}{3H^2}=\frac{6 x+8 y}{3 \left(-4 \beta +(4 \beta +3) x+(4 \beta +3) y-\eta(2 \beta +1)^2 -1\right)}.
\end{equation}

The evolution of the system is then tracked using the e-folding number $ N = \ln(a) $, which serves as a natural time-like variable to describe the expansion of the Universe. The notation prime (') denotes differentiation with respect to $N$, and we obtain an autonomous system of differential equations,

\begin{align}\label{Eq:30}
x' &=x \left(\frac{6 x+8 y}{x(4 \beta +3) +y(4 \beta +3) -\eta(2 \beta +1)^2-4 \beta -1}-3\right),\nonumber\\
y'&=y \left(\frac{6 x+8 y}{x(4 \beta +3) +y(4 \beta +3) -\eta(2 \beta +1)^2-4 \beta -1}-4\right),\nonumber\\
\eta'&=\left(\frac{(3 x+4 y) (x+y-\eta-1)}{x(4 \beta +3) +y(4 \beta +3) -\eta(2 \beta +1)^2-4 \beta -1}\right).
\end{align}

Now, we can obtain the critical points by solving $ x' = 0 $, $ y' = 0 $ and $ \eta' = 0 $. These conditions identify the equilibrium configurations of the system, corresponding to distinct cosmological phases. The nature and stability of each critical point are determined by analyzing the eigenvalues of the Jacobian matrix evaluated at these points. If all three eigenvalues are real and negative, the point is a stable node or attractor, implying that the system evolves naturally toward this configuration in all directions. If all eigenvalues are real and positive, the point is an unstable node, indicating that the system diverges from this state over time. When the eigenvalues include both positive and negative real values, the critical point is classified as a saddle, exhibiting instability along some directions and stability along others. If one or more eigenvalues are complex, the dynamics near the critical point may involve oscillatory or spiral-like behavior. A critical point with complex eigenvalues whose real parts are all negative is referred to as a stable focus, indicating damped oscillations toward equilibrium. Conversely, if the real parts are all positive, the point is an unstable focus. A saddle focus occurs when the eigenvalues include complex components with mixed-sign real parts, signifying that the system is stable in some directions and unstable in others.

 We also express the cosmological parameters, such as the deceleration parameter, equation of state parameter, and density parameters, in terms of the dynamical variables. This enables a physical interpretation of each critical point, allowing us to associate it with specific cosmological epochs, such as matter domination, radiation domination, or accelerated expansion.\\

\begin{align}\label{eq.34}
\Omega_m&=x,~~~\Omega_r=y,~~~\Omega_{f}=1-x-y,\nonumber\\
q&= \left(\frac{3 x+4 y}{x(4 \beta +3) +y(4 \beta +3) -\eta(2 \beta +1)^2-4 \beta -1}-1\right),\nonumber\\
w&= \left(\frac{6 x+8 y}{3 \left(x(4 \beta +3) +y(4 \beta +3) -\eta(2 \beta +1)^2-4 \beta -1\right)}-1\right).
\end{align}

\begin{table}[H]
\renewcommand\arraystretch{0.5}
\centering 
\begin{tabular}{|c| c c c | c | c |} 
\hline\hline 
~~~~~\parbox[c][1.3cm]{2.5cm}{Critical Points
}&~~ $x_c$ ~~& ~~$y_c$~~ &~~ $\eta_c$ ~~& ~~~ Exits for ~~~ & ~~~~Stability~~~~\\ [0.5ex] 
\hline\hline 
\parbox[c][1.3cm]{0.7cm}{$P_f$ } & $0$ & $0$ & $\eta$ &~~~~~ $ 4 \beta+4 \beta^2 \eta+4 \beta \eta+\eta+1\neq 0$ ~~~~~& ~~~~Nonhyperbolic~~~~\\
\hline
\parbox[c][1.3cm]{0.7cm}{$P_m$ }  & $1$ & $0$  & $0$  &  \begin{tabular}{@{}c@{}}Always\end{tabular}  &  Saddle \\
\hline
\parbox[c][1.3cm]{0.7cm}{$P_r$ }  & $0$ & $1$ & $0$ & \begin{tabular}{@{}c@{}}Always\end{tabular}  & Unstable \\
\hline
\end{tabular}
\caption{Critical points, existence conditions, Stability of Critical point corresponding to various phases of the Universe.}
\label{table2}
\end{table}

\begin{figure}
    \centering
    \includegraphics[width=8.5cm]{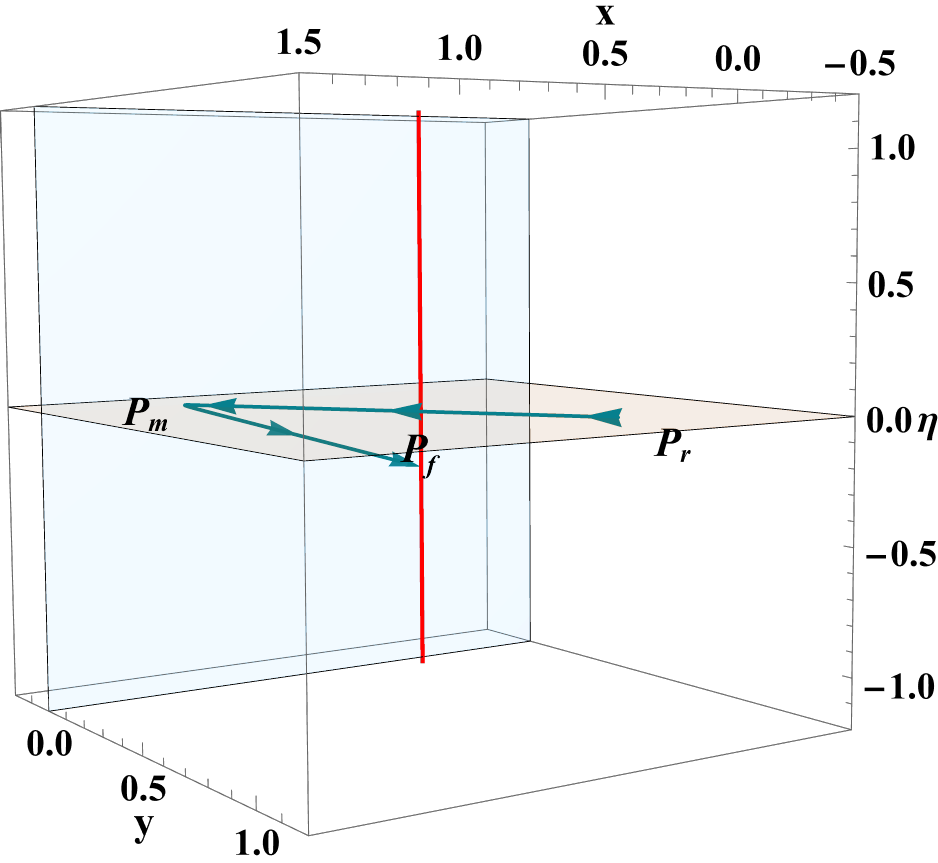}\\
    \includegraphics[width=8.5cm]{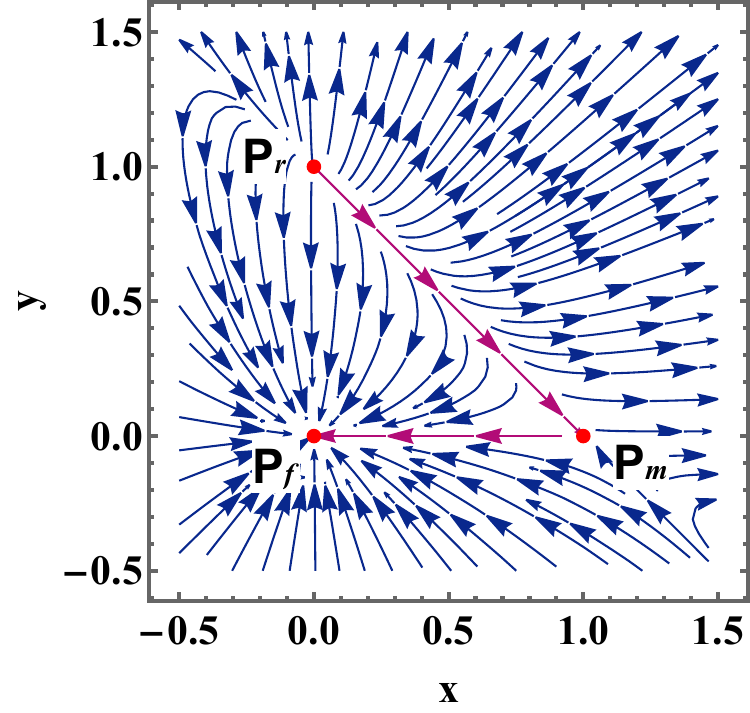}
     \includegraphics[width=8.5cm]{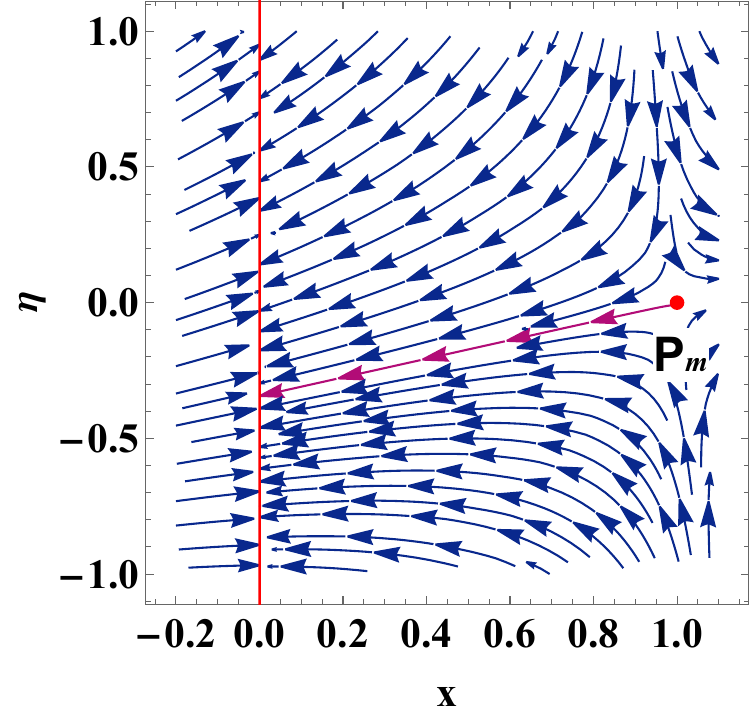}
 \caption{3D Plot (upper panel): Evolution of the Universe in logarithmic $f(Q)$ gravity. The plot highlights the critical points: $\mathbf{P_f}$ (de Sitter accelerated phase), $\mathbf{P_m}$ (matter-dominated phase), and $\mathbf{P_r}$ (radiation-dominated phase). The phase portraits in the $(x, \eta)$-plane show that for the non-hyperbolic critical point $\mathbf{P_f}$, trajectories are attracted along the $\eta$-axis. This behavior is further confirmed in the $(x, y)$-plane, consistently demonstrating the existence of the de Sitter phase.} 
    \label{Fig7}
\end{figure}

\begin{table}[H]
\renewcommand\arraystretch{0.5}
\centering 
\begin{tabular}{|c| c |c| c| c| c|} 
\hline\hline 
~~~~~\parbox[c][1.3cm]{2.5cm}{Critical Points
}& ~~~~$q$~~~~ & ~~~~$\omega$~~~~ & ~~~$\Omega_m$~~~ & ~~~$\Omega_r$~~~ & ~~~$\Omega_{f}$~~~ \\ [0.5ex] 
\hline\hline 
\parbox[c][1.3cm]{0.7cm}{$P_f$ } & $-1$ & $-1$ & $0$ & $0$ & $1$\\
\hline
\parbox[c][1.3cm]{0.7cm}{$P_m$ }  &  $\frac{1}{2}$ & $0$ & $1$ & $0$ & $0$\\
\hline
\parbox[c][1.3cm]{0.7cm}{$P_r$ }  & $1$ & $\frac{1}{3}$ & $0$ & $1$ & $0$\\
\hline
\end{tabular}
\caption{cosmological parameter corresponding to various phases of the Universe.}
\label{table2}
\end{table}
Each critical point has been discussed in detail:

\begin{itemize}

\item \textbf{Critical Points $P_f$  (Dark Energy Phase)}:
This critical point, denoted as $P_f$, is characterized by $x = 0$, $y = 0$, with $\eta$ remaining as a free variable, subject to the existence condition $4 \beta + 4 \beta^2 \eta + 4 \beta \eta + \eta + 1 \neq 0$. This configuration corresponds to the value of deceleration parameter $q = -1$, signifying a de Sitter phase of cosmic evolution. The effective EoS parameter takes the value $w = -1$, reinforcing the interpretation of a Universe dominated by dark energy. The vanishing of the matter and radiation density parameters ($\Omega_m = 0$ and $\Omega_r = 0$) implies that the total energy density is entirely sourced by a dark energy-like component. Specifically, within the framework of $f(Q)$ gravity where the non-metricity scalar $Q$ is a function of the Hubble parameter, this dark energy contribution is quantified by $\Omega_f = 1$. The critical point, denoted $P_f$, yields the eigenvalue set $\left\{0, -4, -3\right\}$ when the Jacobian matrix is evaluated. This combination of negative and zero eigenvalues classifies $P_f$ as a non-hyperbolic critical point. Despite the fact that the point appears to be mathematically unstable, as mentioned in \cite{Coley_1999, aulbach_1984_1058}, the number of vanishing eigenvalues corresponds to the dimensionality of the center manifold. Despite being non-hyperbolic, the eigenvalue structure exhibits normal hyperbolicity, ensuring the attraction of the vector along the z-axis for this critical point. Phase space analysis reveals this attractive nature, particularly for $ \beta = -0.712 $, a value obtained from the observational constraints. By selecting $ \alpha = -0.353 $, consistent with the observationally favored region, the system exhibits a clear tendency to evolve toward this critical point. This behavior is supported by the physical relevance that trajectories near $P_f$ display attracting behavior in the $x\eta$-plane shown in Fig. \ref{Fig7}, consistent with a late-time de Sitter phase in cosmological evolution.\\

\item \textbf{Critical Point $P_m$ (Matter Dominated Phase)}: The critical point $P_m$ represents a significant configuration corresponding to a matter-dominated phase in the evolution of the Universe. Located at $x = 1$, $y = 0$, and $\eta = 0$, it is characterized by the deceleration parameter $q = \frac{1}{2}$ and equation of state $w = 0$. The eigenvalues at this point become
$$
\left\{-1,\ \frac{3}{2} \left(-\sqrt{8 \beta^2 + 4 \beta + 1} - 2 \beta - 1\right),\ \frac{3}{2} \left(\sqrt{8 \beta^2 + 4 \beta + 1} - 2 \beta - 1\right)\right\}.
$$
For the observationally supported value $\beta = -0.712$, this spectrum includes one positive and two negative real eigenvalues. So, it identifies $P_m$ as a saddle point in the phase space. Though the system may approach the point along some directions, it is repelled along others. As a result, this phase is inherently unstable and unable to represent the final state of the Universe. From the physical point of view, the saddle nature of $P_m$ mirrors the transient role of matter in cosmic history. During this epoch, the gravitational influence of matter allowed for the formation of galaxies and large-scale structures. However, as the dynamics unfold, it is ultimately drawn away and transitions into a phase dominated by dark energy. The same behavior has been illustrated in Fig.~\ref{Fig7}, where trajectories in the phase space approach $P_m$ only to diverge later, reflecting the matter domination for a short period of time. Hence, $P_m$ not only encapsulates a mathematical structure but also represents a crucial turning point in the narrative of the Universe.

\item \textbf{Critical Point $P_r$ (Radiation Dominated Phase)}: The critical point $P_r$ corresponds to a radiation-dominated epoch in the cosmological timeline. It is defined by the coordinates $x = 0$, $y = 1$, and $\eta = 0$, representing a phase in which the radiation component entirely dominates the total energy content of the Universe. The associated cosmological parameters are $q = 1$ and $w = \frac{1}{3}$, consistent with a Universe filled with relativistic particles. The radiation density parameter $\Omega_r = 1$ confirms the absence of matter and dark energy contributions in this phase. The eigenvalues at this critical point become,
$$
\left\{1,\ 2 \left(-\sqrt{8 \beta^2 + 4 \beta + 1} - 2 \beta - 1\right),\ 2 \left(\sqrt{8 \beta^2 + 4 \beta + 1} - 2 \beta - 1\right)\right\}.
$$
For $\beta = -0.712$, this spectrum includes two positive and one negative real eigenvalue. Such a configuration identifies the unstable nature of $P_r$. It highlights that the Universe passes through this phase but cannot stay in it indefinitely. The phase portrait shown in Fig.~\ref{Fig7} illustrates this behavior, with all trajectories along $P_r$ diverging. This confirms the early radiation-dominated era of the Universe..

\end{itemize}

\begin{figure}[H]
    \centering
    \includegraphics[width=12cm]{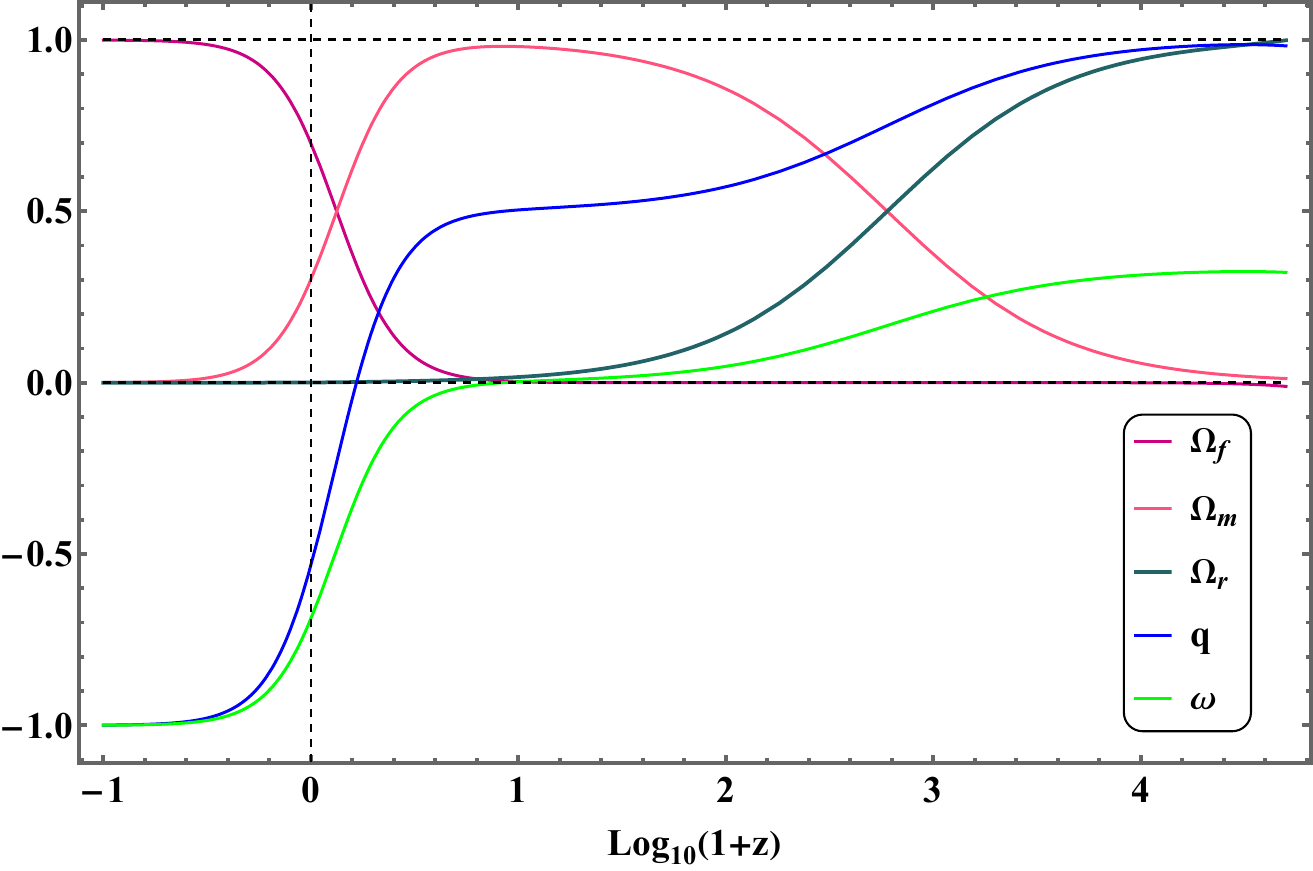}
 \caption{Evolution of Cosmological parameters for Dynamical System analysis using logarithmic $f(Q)$ gravity.} 
    \label{Fig6}
\end{figure}

Fig. \ref{Fig6} illustrates the redshift evolution of the key cosmological quantities in the logarithmic $f(Q,T)$ gravity model, expressed in terms of $\log_{10}(1+z)$ to capture the entire history of the Universe from the radiation era to the present epoch. It shows the fractional energy densities $\Omega_r(z)$ (radiation), $\Omega_m(z)$ (matter), and $\Omega_f(z)$ (effective dark energy), along with the deceleration parameter $q(z)$ and the effective equation-of-state parameter $w(z)$. These quantities are governed by the closure relation

$$
\Omega_r(z) + \Omega_m(z) + \Omega_f(z) = 1,
$$

which ensures that the total energy density matches the critical density, consistent with a spatially flat Universe as supported by CMB \cite{Balkenhol_2023_108} and large-scale structure \cite{Eisenstein_2005_633} observations.

At early times ($z \gg 1$), radiation dominates, with $\Omega_r \approx 1$, reflecting the high-energy density conditions of the primordial Universe. As the Universe expands and cools, matter overtakes radiation, leading to $\Omega_m \approx 1$ during the structure-formation era. At late times, the effective dark energy component becomes dominant ($\Omega_f \to 1$), driving the present cosmic acceleration. The transition between these epochs is clearly visible in the plot. The deceleration parameter $q(z)$ captures this shift: it remains positive during radiation and matter domination (indicating decelerated expansion) but turns negative at low redshift, marking the onset of acceleration with a transition redshift $z_{tr}$ in agreement with observational constraints. The effective equation-of-state parameter $w(z)$ also evolves dynamically, approaching $w \simeq -1$ in the present epoch. This indicates that the model asymptotically mimics a cosmological constant while still allowing deviations at intermediate redshifts, which could leave imprints on observational data.

Physically, this plot encapsulates the full dynamical history of the Universe in the $f(Q,T)$ scenario. It demonstrates that the model consistently reproduces the expected cosmic sequence such as radiation domination, followed by matter domination, and finally dark-energy domination, while maintaining the closure condition throughout cosmic evolution. The evolution of $q(z)$ and $w(z)$ further highlights that the model naturally explains the observed late-time accelerated expansion and provides a geometrical alternative to dark energy.

\section{Results and Discussions}\label{Sec:5} 

In the cosmological observation and dynamical system approaches, the logarithmic form of $f(Q)$ demonstrates its effectiveness in explaining the late-time acceleration of the Universe. Initially, we have transformed the Friedmann equations into a specific form and then solved them numerically using the CC, Pantheon$^+$, and combined datasets. The model parameters are constrained and the values are given in TABLE-\ref{tableA1}. The obtained values resemble the accelerating behavior of the Universe, and in particular, it exhibit quintessence-like behavior. The evolution of Hubble and the distance modulus curve is closely aligned with the $\Lambda$CDM curve.  The state finder diagnostic further confirms that the model exhibits quintessence-like behavior at present and asymptotically approaches $\Lambda$CDM at late times, highlighting the consistency. The age redshift analysis yields $13.913$ Gyr for the $CC$ dataset, $13.671$ Gyr for the Pantheon$^+$ dataset, and $13.527$ Gyr for the CC+Pantheon$^+$ dataset.\\

Also, to examine the late-time stability of cosmic evolution, we performed the dynamical system analysis, reformulating the cosmological equations using dimensionless variables. Here, we have identified and analyzed three critical points, each of which corresponds to key phases in cosmic evolution: the de Sitter, matter-dominated, and radiation-dominated eras. Our analysis reveals that the de Sitter critical point $P_f$ is non-hyperbolic, characterized by a combination of negative and zero eigenvalues. Despite this non-hyperbolicity, the eigenvalue structure exhibits normal hyperbolicity, which guarantees the attraction of trajectories along the $z$-axis for this critical point. The phase space analysis further confirms this attractive nature, indicating that the Universe naturally evolves toward this state. Consequently, the system predicts an epoch of accelerated expansion, in agreement with the currently observed dark energy dominated phase.
 In contrast, the critical point $P_m$ corresponds to the matter-dominated epoch, which exhibits saddle-like instability, where the trajectories approach but do not settle, and the $P_r$ radiation critical points are unstable. These points represent transient decelerating phases. The instability suggests that the Universe evolves away from these phases, transitioning toward acceleration as observed today. Such analytical insights strengthen the physical interpretation of these phases and provide a deeper understanding of how modified gravity, particularly in the logarithmic $f(Q)$ model, can reproduce the complete cosmic history from radiation domination to late-time acceleration.\\

Finally, we have presented a cosmological model of the Universe with some logarithmic form of $f(Q)$ in the framework of $f(Q)$ gravity that exhibits quintessence-like behavior and may provide a compelling alternative to $\Lambda$CDM. The model effectively reproduces key cosmological features, including late-time acceleration, while maintaining dynamical stability and a natural tendency toward a de Sitter phase at late times. These results highlight the significance of exploring modified gravity theories as potential solutions to the accelerated expansion of the Universe.\\

\section*{Acknowledgement} RB acknowledges the financial support provided by the University Grants Commission (UGC) through Junior Research Fellowship UGC-Ref. No.: 211610028858 to carry out the research work. BM acknowledges the support of IUCAA, Pune (India), through the visiting associateship program. The authors are thankful to the esteem referee for the valuable suggestions and comments to improve the quality of the paper.

\section*{References}

\begin{thebibliography}{10}

\bibitem{Riess_1998_116}
{\bf Supernova Search Team} Collaboration, A.~G. Riess {\em et al.}, ``{Observational evidence from supernovae for an accelerating universe and a cosmological constant},'' \href{http://dx.doi.org/10.1086/300499}{{\em Astron. J.} {\bf 116} (1998)  1009--1038}.

\bibitem{Perlmutter_1998_517}
{\bf Supernova Cosmology Project} Collaboration, S.~Perlmutter {\em et al.}, ``{Measurements of $\Omega$ and $\Lambda$ from 42 high redshift supernovae},'' \href{http://dx.doi.org/10.1086/307221}{{\em Astrophys. J.} {\bf 517} (1999)  565--586}.

\bibitem{Giostri_2012_2012_027}
R.~Giostri, M.~V. dos Santos, I.~Waga, {\em et al.}, ``{From cosmic deceleration to acceleration: new constraints from {SN} Ia and {BAO}/{CMB}},'' \href{http://dx.doi.org/10.1088/1475-7516/2012/03/027}{{\em J. Cosmol. Astropart. Phys.} {\bf 2012} (2012) no.~03, 027--027}.

\bibitem{Balkenhol_2023_108}
L.~Balkenhol, D.~Dutcher, A.~Spurio~Mancini, {\em et al.}, ``{Measurement of the CMB temperature power spectrum and constraints on cosmology from the SPT-3G 2018 $TT$, $TE$, and $EE$ dataset},'' \href{http://dx.doi.org/10.1103/PhysRevD.108.023510}{{\em Phys. Rev. D} {\bf 108} (2023) no.~34, 023510}.

\bibitem{Larson_2011_192}
D.~Larson, J.~Dunkley, G.~Hinshaw, {\em et al.}, ``{SEVEN-YEAR WILKINSON MICROWAVE ANISOTROPY PROBE (WMAP) OBSERVATIONS: POWER SPECTRA AND WMAP-DERIVED PARAMETERS},'' \href{http://dx.doi.org/10.1088/0067-0049/192/2/16}{{\em The Astrophysical Journal Supplement Series} {\bf 192} (2011) no.~2, 16}.

\bibitem{Hinshaw_2013_208}
G.~Hinshaw, D.~Larson, E.~Komatsu, {\em et al.}, ``{Nine-year Wilkison Microwave Anisotropy Probe (WMAP) Observations: Cosmological Parameter Results},'' \href{http://dx.doi.org/10.1088/0067-0049/208/2/19}{{\em Astrophys. J. Supp. Ser.} {\bf 208} (2013)  19}.

\bibitem{Bennett_2003_148}
C.~L. Bennett, M.~Halpern, G.~Hinshaw, {\em et al.}, ``First-year wilkinson microwave anisotropy probe (wmap)* observations: Preliminary maps and basic results,'' \href{http://dx.doi.org/10.1086/377253}{{\em Astrophys. J. Supp. Ser.} {\bf 148} (2003) no.~1, 1}.

\bibitem{Spergel_2003_148}
D.~N. Spergel {\em et al.}, ``First year wilkinson microwave anisotropy probe ({WMAP}) observations: Determination of cosmological parameters,'' \href{http://dx.doi.org/10.1086/377226}{{\em Astrophys. J. Supp. Ser.} {\bf 148} (2003) no.~1, 175--194}.

\bibitem{Ade_2016_594a}
P.~A.~R. Ade, N.~Aghanim, M.~Arnaud, M.~Ashdown, {\em et al.}, ``{Planck 2015 results},'' \href{http://dx.doi.org/10.1051/0004-6361/201525830}{{\em {Astronomy $\&$ Astrophysics}} {\bf 594} (2016)  A13}.

\bibitem{Aghanim_2018_641}
{\bf Planck} Collaboration, N.~Aghanim, Y.~Akrami, M.~Ashdown, {\em et al.}, ``{Planck 2018 results. VI. Cosmological parameters},'' \href{http://dx.doi.org/10.1051/0004-6361/201833910}{{\em Astron. Astrophys.} {\bf 641} (2020)  A6}. [Erratum: Astron.Astrophys. 652, C4 (2021)].

\bibitem{Alam_2017_470}
S.~Alam, M.~Ata, S.~Bailey, F.~Beutler, {\em et al.}, ``{The clustering of galaxies in the completed SDSS-III Baryon Oscillation Spectroscopic Survey: cosmological analysis of the DR12 galaxy sample},'' \href{http://dx.doi.org/10.1093/mnras/stx721}{{\em Monthly Notices of the Royal Astronomical Society} {\bf 470} (2017) no.~3, 2617--2652}.

\bibitem{Eisenstein_2005_633}
D.~J. Eisenstein, I.~Zehavi, D.~W. Hogg, R.~Scoccimarro, {\em et al.}, ``{Detection of the Baryon Acoustic Peak in the Large-Scale Correlation Function of SDSS Luminous Red Galaxies},'' \href{http://dx.doi.org/10.1086/466512}{{\em The Astrophysical Journal} {\bf 633} (2005) no.~2, 560}.

\bibitem{Abbott_2018_480}
{\bf DES} Collaboration, T.~M.~C. Abbott {\em et al.}, ``{Dark Energy Survey Year 1 Results: A Precise H0 Estimate from DES Y1, BAO, and D/H Data},'' \href{http://dx.doi.org/10.1093/mnras/sty1939}{{\em Mon. Not. Roy. Astron. Soc.} {\bf 480} (2018) no.~3, 3879--3888}.

\bibitem{Abbott_2016_460}
D.~E.~S. Collaboration:, T.~Abbott, F.~B. Abdalla, J.~Aleksi\'c', {\em et al.}, ``{The Dark Energy Survey: more than dark energy $-$ an overview},'' \href{http://dx.doi.org/10.1093/mnras/stw641}{{\em Monthly Notices of the Royal Astronomical Society} {\bf 460} (2016) no.~2, 1270--1299}.

\bibitem{Riess_2019_876}
A.~G. Riess, S.~Casertano, W.~Yuan, {\em et al.}, ``{Large Magellanic Cloud Cepheid Standards Provide a 1\% Foundation for the Determination of the Hubble Constant and Stronger Evidence for Physics beyond $\Lambda$CDM},'' \href{http://dx.doi.org/10.3847/1538-4357/ab1422}{{\em Astrophys. J.} {\bf 876} (2019) no.~1, 85}.

\bibitem{Raichoor_2020_500}
A.~Raichoor, A.~de~Mattia, A.~J. Ross, {\em et al.}, ``{The completed {SDSS}-{IV} extended Baryon Oscillation Spectroscopic Survey: large-scale structure catalogues and measurement of the isotropic {BAO} between redshift 0.6 and 1.1 for the Emission Line Galaxy Sample},'' \href{http://dx.doi.org/10.1093/mnras/staa3336}{{\em Mon. Not. Roy. Astron. Soc.} {\bf 500} (2020) no.~3, 3254--3274}.

\bibitem{Cai_2005_2005_002}
R.-G. Cai and A.~Wang, ``{Cosmology with interaction between phantom dark energy and dark matter and the coincidence problem},'' \href{http://dx.doi.org/10.1088/1475-7516/2005/03/002}{{\em J. Cosmol. Astropart. Phys.} {\bf 2005} (2005) no.~-, 002}.

\bibitem{Weinberg_1989_61}
S.~Weinberg, ``{The Cosmological Constant Problem},'' \href{http://dx.doi.org/10.1103/RevModPhys.61.1}{{\em Rev. Mod. Phys.} {\bf 61} (1989)  1--23}.

\bibitem{Lobo_2009_173}
F.~S.~N. Lobo, ``The dark side of gravity: Modified theories of gravity,'' \href{http://dx.doi.org/10.48550/ARXIV.0807.1640}{{\em Dark Energy-Current Advances and Ideas} (2009)  173--204}.

\bibitem{Carroll_1998_81}
S.~M. Carroll, ``{Quintessence and the Rest of the World: Suppressing Long-Range Interactions},'' \href{http://dx.doi.org/10.1103/PhysRevLett.81.3067}{{\em Phys. Rev. Lett.} {\bf 81} (1998) no.~0, 3067--3070}.

\bibitem{Bhagat_2023_42}
R.~Bhagat, S.~Narawade, B.~Mishra, and S.~Tripathy, ``{Constrained cosmological model in f(Q,T) gravity with non-linear non-metricity},'' \href{http://dx.doi.org/10.1016/j.dark.2023.101358}{{\em Physics of the Dark Universe} {\bf 42} (2023) no.~-, 101358}.

\bibitem{Picon_2001_63_103510}
C.~Armendariz-Picon, V.~Mukhanov, and P.~J. Steinhardt, ``{Essentials of k-essence},'' \href{http://dx.doi.org/10.1103/PhysRevD.63.103510}{{\em Phys. Rev. D} {\bf 63} (2001)  103510}.

\bibitem{Odintsov_2018_98}
S.~D. Odintsov and V.~K. Oikonomou, ``{Dynamical systems perspective of cosmological finite-time singularities in $f\mathbf{(}R\mathbf{)}$ gravity and interacting multifluid cosmology},'' \href{http://dx.doi.org/10.1103/PhysRevD.98.024013}{{\em Phys. Rev. D} {\bf 98} (2018) no.~25, 024013}.

\bibitem{Hrycyna_2007_76}
O.~Hrycyna and M.~Szyd\l{}owski, ``Extended quintessence with nonminimally coupled phantom scalar field,'' \href{http://dx.doi.org/10.1103/PhysRevD.76.123510}{{\em Phys. Rev. D} {\bf 76} (2007)  123510}.

\bibitem{Jimenez_2018_98}
J.~B. Jim\'enez, L.~Heisenberg, and T.~Koivisto, ``{Coincident general relativity},'' \href{http://dx.doi.org/10.1103/PhysRevD.98.044048}{{\em Phys. Rev. D} {\bf 98} (2018) no.~6, 044048}.

\bibitem{Cai_2016_79}
Y.-F. Cai, S.~Capozziello, M.~D. Laurentis, and E.~N. Saridakis, ``${f(T)}$ teleparallel gravity and cosmology,'' \href{http://dx.doi.org/10.1088/0034-4885/79/10/106901}{{\em Reports on Progress in Physics} {\bf 79} (2016) no.~10, 106901}.

\bibitem{Jimenez_2020_101}
J.~B. Jim\'enez, L.~Heisenberg, T.~Koivisto, and S.~Pekar, ``{Cosmology in $f(Q)$ geometry},'' \href{http://dx.doi.org/10.1103/PhysRevD.101.103507}{{\em Phys. Rev. D} {\bf 101} (2020) no.~16, 103507}.

\bibitem{Narwade_2025_409}
S.~Narawade and B.~Mishra, ``{Insights into f(Q) gravity: Modeling through deceleration parameter},'' \href{http://dx.doi.org/10.1016/j.jheap.2025.01.013}{{\em Journal of High Energy Astrophysics} {\bf 45} (2025) no.~-, 409--417}.

\bibitem{HEISENBERG20241}
L.~Heisenberg, ``Review on f(q) gravity,'' \href{http://dx.doi.org/https://doi.org/10.1016/j.physrep.2024.02.001}{{\em Physics Reports} {\bf 1066} (2024)  1--78}.

\bibitem{Dimakis_2025_273}
N.~Dimakis, P.~A. Terzis, A.~Paliathanasis, and T.~Christodoulakis, ``{Static, spherically symmetric solutions in f(Q)-gravity and in nonmetricity scalar-tensor theory},'' \href{http://dx.doi.org/10.1016/j.jheap.2024.12.011}{{\em Journal of High Energy Astrophysics} {\bf 45} (2025) no.~-, 273--289}.

\bibitem{Vignolo_2024}
S.~Vignolo, F.~Esposito, and S.~Carloni, ``{A note on the junction conditions in f({\cal Q})-gravity},'' \href{http://dx.doi.org/10.1088/1361-6382/ad6be0}{{\em Classical and Quantum Gravity} {\bf 41} (2024) no.~18, 187001}.

\bibitem{Paliathanasis_2024_46}
A.~Paliathanasis, ``{Dipole cosmology in f(Q) gravity},'' \href{http://dx.doi.org/10.1016/j.dark.2024.101585}{{\em Physics of the Dark Universe} {\bf 46} (2024) no.~-, 101585}.

\bibitem{Alwan_2024}
M.~A. Alwan, T.~Inagaki, B.~Mishra, and S.~Narawade, ``{Neutron star in covariant f(Q) gravity},'' \href{http://dx.doi.org/10.1088/1475-7516/2024/09/011}{{\em Journal of Cosmology and Astroparticle Physics} {\bf 2024} (2024) no.~09, 011}.

\bibitem{Jensko_2025}
E.~Jensko, ``{Spatial curvature in coincident gauge f(Q) cosmology},'' \href{http://dx.doi.org/10.1088/1361-6382/adadbf}{{\em Classical and Quantum Gravity} {\bf 42} (2025) no.~5, 055011}.

\bibitem{Chen_2025}
W.-X. Chen, ``{Calculating the Hawking temperature of black holes in f(Q) gravity using the RVB method: a residue-based approach},'' \href{http://dx.doi.org/10.1139/cjp-2024-0214}{{\em Canadian Journal of Physics} {\bf -} (2025) no.~-, --}.

\bibitem{Rastgoo_2024_84}
S.~Rastgoo and F.~Parsaei, ``{Traversable wormholes satisfying energy conditions in f(Q) gravity},'' \href{http://dx.doi.org/10.1140/epjc/s10052-024-12939-8}{{\em The European Physical Journal C} {\bf 84} (2024) no.~-, 563}.

\bibitem{Agrawal_2023_83}
A.~Agrawal, B.~Mishra, and P.~Agrawal, ``{Matter bounce scenario in extended symmetric teleparallel gravity},'' \href{http://dx.doi.org/10.1140/epjc/s10052-023-11266-8}{{\em The European Physical Journal C} {\bf 83} (2023) no.~2, 113}.

\bibitem{Capozziello_2022_832}
S.~Capozziello and R.~DAgostino, ``{Model$-$independent reconstruction of $f(Q)$ non$-$metric gravity},'' \href{http://dx.doi.org/10.1016/j.physletb.2022.137229}{{\em Physics Letters B} {\bf 832} (2022) no.~-, 137229}.

\bibitem{Najera_2023_524}
J.~A. Najera, C.~Araoz~Alvarado, and C.~Escamilla-Rivera, ``{Constraints on f(Q) logarithmic model using gravitational wave standard sirens},'' \href{http://dx.doi.org/10.1093/mnras/stad2180}{{\em Monthly Notices of the Royal Astronomical Society} {\bf 524} (2023) no.~4, 5280--5290}.

\bibitem{Ambrosio_2022_105}
F.~D'Ambrosio, S.~D.~B. Fell, L.~Heisenberg, and S.~Kuhn, ``{Black holes in $f(Q)$ gravity},'' \href{http://dx.doi.org/10.1103/PhysRevD.105.024042}{{\em Phys. Rev. D} {\bf 105} (2022) no.~35, 024042}.

\bibitem{BHAGAT2025101913}
R.~Bhagat, F.~Tello-Ortiz, and B.~Mishra, ``{Tracing cosmic evolution through Weyl-Type f(Q,T) gravity model: Theoretical analysis and observational validation},'' \href{http://dx.doi.org/10.1016/j.dark.2025.101913}{{\em Physics of the Dark Universe} {\bf 48} (2025)  101913}.

\bibitem{Moresco_2015_450}
M.~Moresco, ``{Raising the bar: new constraints on the Hubble parameter with cosmic chronometers at z~$\sim$~2},'' \href{http://dx.doi.org/10.1093/mnrasl/slv037}{{\em Mon. Not. Roy. Astron. Soc.: Lett.} {\bf 450} (2015) no.~1, L16--L20}.

\bibitem{Scolnic_2022}
D.~Scolnic, D.~Brout, A.~Carr, A.~G. Riess, {\em et al.}, ``The pantheon+ analysis: The full data set and light-curve release,'' \href{http://dx.doi.org/10.3847/1538-4357/ac8b7a}{{\em The Astrophysical Journal} {\bf 938} (2022) no.~2, }.

\bibitem{Anagnostopoulos_2021_822}
F.~K. Anagnostopoulos, S.~Basilakos, and E.~N. Saridakis, ``{First evidence that non-metricity $f(Q)$ gravity could challenge $\Lambda$CDM},'' \href{http://dx.doi.org/10.1016/j.physletb.2021.136634}{{\em Physics Letters B} {\bf 822} (2021)  136634}.

\bibitem{SULTANA2025100422}
S.~Sultana and S.~Chattopadhyay, ``Constraining exponential f(q) gravity with cosmic chronometers and supernovae: A data-driven analysis,'' \href{http://dx.doi.org/https://doi.org/10.1016/j.jheap.2025.100422}{{\em Journal of High Energy Astrophysics} {\bf 48} (2025)  100422}.

\bibitem{B_hmer_2023}
Böhmer, Christian and Jensko, Erik and Lazkoz, Ruth, ``Dynamical Systems Analysis of f(Q) Gravity,'' \href{http://dx.doi.org/10.3390/universe9040166}{{\em Universe} {\bf 9} (2023)  4}.

\bibitem{PALIATHANASIS2025101993}
A.~Paliathanasis, ``Testing non-coincident f(q)-gravity with desi dr2 bao and grbs,'' \href{http://dx.doi.org/https://doi.org/10.1016/j.dark.2025.101993}{{\em Physics of the Dark Universe} {\bf 49} (2025)  101993}.

\bibitem{Foreman-Mackey_2013_125}
D.~Foreman-Mackey, D.~W. Hogg, D.~Lang, and J.~Goodman, ``{emcee: The MCMC Hammer},'' \href{http://dx.doi.org/10.1086/670067}{{\em Publications of the Astronomical Society of the Pacific} {\bf 125} (2013) no.~925, 306}.

\bibitem{Moresco_2022_25}
M.~Moresco, L.~Amati, L.~Amendola, {\em et al.}, ``{Unveiling the Universe with emerging cosmological probes},'' \href{http://dx.doi.org/10.1007/s41114-022-00040-z}{{\em Living Rev. Rel.} {\bf 25} (2022)  6}.

\bibitem{Brout_2022_938}
D.~Brout, D.~Scolnic, B.~Popovic, {\em et al.}, ``{The Pantheon+ Analysis: Cosmological Constraints},'' \href{http://dx.doi.org/10.3847/1538-4357/ac8e04}{{\em Astrophys. J.} {\bf 938} (2022) no.~2, 110}.

\bibitem{Sahni_2003_77}
V.~Sahni, T.~Saini, A.~Starobinsky, and U.~Alam, ``{Statefinder--A new geometrical diagnostic of dark energy},'' \href{http://dx.doi.org/10.1134/1.1574831}{{\em JETP Letters} {\bf 77} (2003) no.~5, 201 -- 206}.

\bibitem{Alam_2003}
U.~Alam, V.~Sahni, T.~Deep~Saini, and A.~A. Starobinsky, ``{Exploring the expanding Universe and dark energy using the statefinder diagnostic},'' \href{http://dx.doi.org/10.1046/j.1365-8711.2003.06871.x}{{\em Monthly Notices of the Royal Astronomical Society} {\bf 344} (2003) no.~4, 1057}.

\bibitem{Age_Hubble}
``Hubble finds birth certificate of oldest known star,'' {\em Phys.Org. 2013-03-07}  .

\bibitem{Plank_2020}
N.~Aghanim, Y.~Akrami, {\em et al.}, ``{Planck 2018 results},'' \href{http://dx.doi.org/10.1051/0004-6361/201833910}{{\em Astronomy and Astrophysics} {\bf 641} (2020)  67}.

\bibitem{Creevey_2015}
O.~L. Creevey, F.~Thevenin, P.~Berio, U.~Heiter, {\em et al.}, ``Benchmark stars for gaia fundamental properties of the population ii star hd 140283 from interferometric, spectroscopic, and photometric data,'' \href{http://dx.doi.org/10.1051/0004-6361/201424310}{{\em Astronomy and amp} {\bf 575} (2015)  A26}.

\bibitem{Tang_2021}
J.~Tang and M.~Joyce, ``Revised best estimates for the age and mass of the methuselah star hd 140283 using mesa and interferometry and implications for 1d convection,'' \href{http://dx.doi.org/10.3847/2515-5172/ac01ca}{{\em Research Notes of the AAS} {\bf 5} (2021) no.~5, 117}.

\bibitem{Cowan_2002}
J.~J. Cowan, C.~Sneden, S.~Burles, I.~I. Ivans, T.~C. Beers, J.~W. Truran, J.~E. Lawler, F.~Primas, G.~M. Fuller, B.~Pfeiffer, and K.~Kratz, ``{The Chemical Composition and Age of the Metal$-$poor Halo Star BD $+17o3248$},'' \href{http://dx.doi.org/10.1086/340347}{{\em The Astrophysical Journal} {\bf 572} (2002)  861}.

\bibitem{Paliathanasis_2023_41}
A.~Paliathanasis, ``{Dynamical analysis of f(Q) cosmology},'' \href{http://dx.doi.org/10.1016/j.dark.2023.101255}{{\em Physics of the Dark Universe} {\bf 41} (2023) no.~-, 101255}.

\bibitem{Bhagat_ASPdyna2024}
R.~Bhagat and B.~Mishra, ``{Observational constrained Weyl type f(Q,T) gravity cosmological model and the dynamical system analysis},'' \href{http://dx.doi.org/10.1016/j.astropartphys.2024.103011}{{\em Astroparticle Physics} {\bf 163} (2024) no.~-, 103011}.

\bibitem{Bohmer_2016_book_dyna}
C.~G. B\"{o}hmer and N.~Chan, ``{Dynamical Systems in Cosmology},'' \href{http://dx.doi.org/10.1142/9781786341044_0004}{{\em Dynamical and Complex Systems} {\bf -} (2017) no.~-, 121--156}.


\bibitem{B_hmer_2022}
C.~G. B\"{o}hmer  and Jensko, Erik and Lazkoz, Ruth, ``{Cosmological dynamical systems in modified gravity},'' \href{http://dx.doi.org/10.1140/epjc/s10052-022-10412-y}{{\em The European Physical Journal C} {\bf -} (2022) no.~-, 6}.

\bibitem{Coley_1999}
A.~A. Coley, ``{Dynamical systems in cosmology},'' in {\em Spanish Relativity Meeting (ERE 99)}.
\newblock 9, 1999.
\newblock \href{http://arxiv.org/abs/gr-qc/9910074}{{\tt arXiv:gr-qc/9910074}}.

\bibitem{aulbach_1984_1058}
B.~Aulbach, ``Continuous and discrete dynamics near manifolds of equilibria,'' {\em Lecture notes in mathematics} {\bf 1058} (1984)  .

 \end{thebibliography}


\end{document}